\documentclass[prd,aps,showpacs,11pt,
nofootinbib,%
tightenlines,
preprintnumbers,eqsecnum%
]{revtex4-1}

\usepackage{graphicx}
\usepackage{amsmath}
\usepackage{amssymb}
\usepackage{hyperref} 

\newcommand{\be}{\begin{equation}}
\newcommand{\ee}{\end{equation}}
\newcommand{\bea}{\begin{eqnarray}}
\newcommand{\eea}{\end{eqnarray}}

\newcommand{\6}{\partial }

\newcommand{\R}{r_{\rm KK}}
\newcommand{\RSS}{R_{\rm D4}}
\newcommand{\MKK}{M_{\rm KK}}
\newcommand{\UKK}{U_{\rm KK}}
\newcommand{\gYM}{g_{\rm YM}}
\newcommand{\Nc}{N_{c}}
\newcommand{\ls}{l_{s}}
\newcommand{\gs}{g_{s}}

\newcommand{\Tr}{{\rm Tr}\,}

\newcommand{\D}{G_{D}}
\newcommand{\Ds}{G_{D}^*}
\newcommand{\G}{G_{E}}
\newcommand{\Gs}{G_{E}^*}
\newcommand{\HG}{H_{E}}
\newcommand{\MG}{M_{E}}
\newcommand{\CG}{\mathcal C_{E}}
\newcommand{\mr}{m_{\rho}}

\def\PDG{\cite{Agashe:2014kda}}
\begin{document}

\title{Glueball Decay Rates 
in the Witten-Sakai-Sugimoto Model}

\preprint{TUW-15-01}

\author{Frederic Br\"unner}
\author{Denis Parganlija}
\author{Anton Rebhan}
\affiliation{Institut f\"ur Theoretische Physik, Technische Universit\"at Wien,
        Wiedner Hauptstrasse 8-10, A-1040 Vienna, Austria}

\date{\today}

\begin{abstract}
We revisit and extend previous calculations of 
glueball decay rates in the Sakai-Sugimoto model,
a holographic top-down approach for QCD with chiral quarks based on D8-$\overline{\rm D8}$ 
probe branes in Witten's holographic model of 
nonsupersymmetric Yang-Mills theory.
The rates for decays into two pions, two vector mesons,
four pions, and the strongly suppressed decay into four $\pi^0$
are worked out quantitatively, using a range of the 't Hooft coupling
which closely reproduces the decay rate of $\rho$ and $\omega$ mesons
and also leads to a gluon condensate consistent with QCD sum rule calculations.
The lowest holographic glueball, which arises
from a rather exotic polarization of gravitons in the supergravity
background, turns out to have a significantly lower mass
and larger width than the
two widely discussed glueball candidates $f_0(1500)$ and $f_0(1710)$.
The lowest
nonexotic and predominantly dilatonic scalar
mode, which has a mass of 1487 MeV in the Witten-Sakai-Sugimoto model, instead provides
a narrow glueball state, and we conjecture that only this nonexotic
mode should be identified with a scalar glueball
component of $f_0(1500)$ or $f_0(1710)$.
Moreover the decay pattern of the tensor glueball is determined, which
is found to have a comparatively broad total width when its mass is
adjusted to around or above 2 GeV.
\end{abstract}

\maketitle

\tableofcontents

\section{Introduction}

The nonabelian nature of Quantum Chromodynamics (QCD)---the theory of the strong interactions---makes it possible to form bound states of gauge bosons, the so-called glueballs \cite{Fritzsch:1972jv,Fritzsch:1975tx,Jaffe:1975fd}.
In pure Yang-Mills theory, these are in fact the only possible particle states
and their spectrum has been studied
in detail in lattice gauge theory \cite{Morningstar:1999rf,Chen:2005mg,Loan:2005ff}.
Glueballs are obtained for a range of quantum numbers $J^{PC}$,
where $J$ denotes total spin, $P$ parity, and $C$ charge conjugation; the lowest
glueball has the quantum numbers of the vacuum, $J^{PC}=0^{++}$.
In the presence of quarks, the situation becomes complicated because
glueballs can mix with $q\bar q$ states of the same quantum numbers.
Lattice simulations of QCD including quarks are more difficult, but
recent unquenched calculations
continue to indicate the existence of glueballs \cite{Gregory:2012hu}
with the lightest glueball around 1600--1800 MeV.

The identification of glueballs in experimental data, however, remains elusive
\cite{Bugg:2004xu,Klempt:2007cp,Crede:2008vw,Ochs:2013gi}
and will be one of the objectives of the PANDA experiment at FAIR \cite{Lutz:2009ff,Wiedner:2011mf}.
Experimentally, the $0^{++}$ meson sector turns out to be
particularly challenging. 
Listings of the Particle Data Group (PDG) \PDG\ contain five isospin-zero scalar states in 
the energy region below 2 GeV: $f_0(500)$ or $\sigma$,  $f_0(980)$,  $f_0(1370)$,  $f_0(1500)$ and  $f_0(1710)$,
with the last two rather narrow states being frequently discussed as
potential candidates for states with dominant glueball content
\cite{Amsler:1995td,Lee:1999kv,Close:2001ga,Amsler:2004ps,Close:2005vf,Giacosa:2005zt,Albaladejo:2008qa,Mathieu:2008me,Janowski:2011gt,Janowski:2014ppa}.
Alternative scenarios with broad glueball resonances around 1 GeV and
mixing with the $\sigma$ meson are also discussed
in the literature \cite{Narison:1996fm,Minkowski:1998mf,Minkowski:2002nf,Ochs:2006rb}.


A similarly unclear situation is found 
in the case of the lightest tensor glueball: lattice simulations obtain a
mass between 2.3 GeV and 2.6 GeV \cite{Morningstar:1999rf,Chen:2005mg,Loan:2005ff,Gregory:2012hu}
while the PDG lists $f_2(1950)$, $f_2(2010)$, $f_2(2300)$ and $f_2(2340)$ as established states
around and above 2 GeV, with several needing confirmation 
[e.g., the narrow $f_J(2220)$ state that may have spin two or four, or may not exist at all].
Various approaches to low-energy QCD have also been applied to this region \cite{Burakovsky:1999ug,Cotanch:2005ja,Anisovich:2004vj,Anisovich:2005iv,
Giacosa:2005bw} but a clear identification of a tensor glueball in the
meson spectrum is missing.

A central difficulty is the paucity of theoretical predictions of glueball couplings and
decay rates from first principles. Lattice gauge theory provides information on Euclidean
correlators and the extraction of real-time quantities is involved and fraught with uncertainties.
Glueballs are particularly difficult to study when dynamical quarks are included.

A completely different approach to strongly coupled gauge theories has been developed over the last
one and a half decades in the form of anti-de Sitter/conformal field theory (AdS/CFT) correspondence, 
or more generally, gauge-string duality \cite{Maldacena:1997re,Aharony:1999ti}.
The AdS/CFT correspondence posits a map of correlation functions of
gauge invariant composite operators with large number of colors $N_c$ and large 't Hooft coupling
to perturbations of certain backgrounds in classical (super-) gravity.
Already in 1998, Witten \cite{Witten:1998zw} proposed a top-down construction of such a duality based
on type-IIA supergravity, where both supersymmetry and conformal invariance are broken
such 
that at low energies, below a Kaluza-Klein mass scale $\MKK$,
the dual gauge theory is four-dimensional large-$N_c$ Yang-Mills theory.
The calculation of glueball spectra from type-IIA supergravity was in fact one of the first
applications of ``holographic QCD'' \cite{hep-th/9805129,Csaki:1998qr,
Hashimoto:1998if,Csaki:1999vb,Constable:1999gb,Brower:2000rp}. (Glueballs have 
subsequently been studied further in more phenomenological, bottom-up holographic models
in, e.g., Ref.~\cite{BoschiFilho:2002ta,Colangelo:2007pt,Forkel:2007ru}.)

Quarks in the fundamental representation can be added to the AdS/CFT correspondence
in the form of probe flavor D-branes \cite{Karch:2002sh}. In type-IIA superstring theory there
are D-branes of even spatial dimensionality, and the first attempt to include quarks
in Witten's model of nonsupersymmetric Yang-Mills theory was based on D6 branes
\cite{Kruczenski:2003uq}. This made it possible to study 
chiral symmetry breaking in the case
of one flavor, which however did not permit a correct generalization to
flavor number $N_f>1$, an issue that was solved in 2004 by Sakai and Sugimoto \cite{Sakai:2004cn,Sakai:2005yt} by adding pairs of D8 and anti-D8 branes
intersecting the color D4 branes of the Witten model. This model
has been remarkably successful in reproducing various features of
low-energy QCD while being firmly rooted in string theory with a minimal
set of parameters---for given $N_c$ and $N_f$, the only dimensionless parameter is
the 't Hooft coupling $\lambda$ at the Kaluza-Klein scale $\MKK$.

In this paper we shall use the Witten-Sakai-Sugimoto model to study
glueball-meson interactions and to calculate glueball decay rates from
the resulting effective interaction Lagrangians. This was first
carried out by Hashimoto, Tan, and Terashima in Ref.~\cite{Hashimoto:2007ze},
whose calculations we repeat (with important corrections) and extend. 

In addition to the lowest glueball mode in the Witten model, which happens
to be rather different from the dilaton mode that plays this role in
simpler bottom-up models of holographic QCD, we consider the (predominantly
but not purely) dilatonic mode of the Witten model, as well as the tensor glueball
and their excitations. We calculate decay rates into two and four pions, and
we confirm the prediction of Ref.~\cite{Hashimoto:2007ze} that scalar glueball
decay into four $\pi^0$ mesons is suppressed by evaluating the rate
quantitatively. The latter receives contributions from multi-glueball interactions
as well as from higher-order terms in the DBI action of the D8 branes,
with the latter yielding the dominant piece.

One of the main conclusion of our work is that the lowest gravitational mode
in the Witten-Sakai-Sugimoto model appears to be ill suited to model
the lowest glueball of QCD as found in
lattice simulations, while the dilatonic mode has reasonable properties
regarding its mass and decay rates. 
The lowest mode either has to be discarded on grounds of
its exotic polarization along the compactified dimension of the type-IIA background
or perhaps could find a physical role as a pure-glue component of the $\sigma$-meson
\cite{Narison:1996fm} (which itself
is absent in the Sakai-Sugimoto model) or the ``red dragon'' of Ref.~\cite{Minkowski:1998mf}.

We also make quantitative comparisons
with experimental data on glueball candidates among scalar mesons at or
above 1.5 GeV by extrapolating the mass of the holographic glueball
and assuming weak mixing with $q\bar q$ states as the latter
is parametrically suppressed at large $N_c$ \cite{Lucini:2012gg}
and thus also in the Witten-Sakai-Sugimoto model \cite{Hashimoto:2007ze}. 
Moreover, the decay pattern of the tensor glueball is worked out
in detail, where also extrapolations to decays into massive pseudo-Goldstone bosons
appear possible.

In view of Refs.~\cite{Ellis:1984jv,Janowski:2014ppa},
a particularly interesting feature of the holographic approach is that it admits
narrow glueball states in the mass range predicted by
lattice simulations, while the prediction of the gluon condensate
is small, close to its standard SVZ value \cite{Shifman:1978bx}.

\section{The Witten model of nonsupersymmetric Yang-Mills theory}

\begin{table}
\begin{tabular}{c|ccc}
\toprule
Here & \cite{Hashimoto:2007ze} & \cite{Brower:2000rp} & \cite{Sakai:2004cn} \\
\colrule
$x^{11}$ & $x^4$ & $x^{11}$ & -- \\
$R_{11}$ & $R_{11}$ & $R_1$ & -- \\
$x^4$ & $\tau$ & $\tau$ & $\tau$ \\ 
$R_4\equiv \MKK^{-1}$ &  $\MKK^{-1}$ & $R_2$ & $\MKK^{-1}$\\
$\R$ & $R$ & $R$ & -- \\
$\RSS\equiv L/2$ & $R_{\rm SS}$ & $R_{\rm AdS}/2$ & $R$ \\
\botrule
\end{tabular}
\caption{
Notations used here versus notation in Hashimoto et al.\ \cite{Hashimoto:2007ze},
Brower et al.\ \cite{Brower:2000rp},
and Sakai \& Sugimoto \cite{Sakai:2004cn,Sakai:2005yt}
}
\label{tabnotation}
\end{table}

The Witten model of nonsupersymmetric (and nonconformal) Yang-Mills theory in
3+1 dimensions is based on 
the AdS/CFT correspondence for a six-dimensional (0,2) superconformal field theory
that is obtained from a large number $N_c$
of coincident M5 branes in 11-dimensional M-theory. 
Their near-horizon 11-d supergravity geometry is the product space AdS$_7\times S^4$
with a curvature radius $L$ of the AdS$_7$ space that is twice the radius of the $S^4$.
With M5 branes extended along the directions 0,1,2,3,4, and 11,
the line element of this space reads \cite{Becker:2007zj}
\be
ds^2=\frac{r^2}{L^2}\left[ \eta_{\mu\nu}dx^\mu dx^\nu+(dx^4)^2+(dx^{11})^2
 \right]+\frac{L^2}{r^2}{dr^2}+\frac{L^2}{4}d\Omega_4^2,
\ee
where $\mu,\nu=0,\ldots,3$ are (3+1)-dimensional indices (following \cite{Becker:2007zj} we are skipping
the index value 10).
The six-dimensional gauge theory living on the boundary of AdS$_7$ is
a rather elusive maximally supersymmetric conformal field theory without a
Lagrangian formulation.
Dimensional reduction on a supersymmetry preserving circle with
\be
x^{11}\simeq x^{11}+2\pi R_{11},\quad  R_{11}=\gs\ls,\quad \ls^2=\alpha'
\ee
leads to the near-horizon geometry of (nonconformal) D4 branes of type-IIA supergravity,
whose dual theory is a five-dimensional super-Yang-Mills theory.

Already in 1998, Witten proposed to use this correspondence as a basis for a
holographic model of the low-energy regime of pure-glue Yang-Mills theory
by a further circle compactification which breaks supersymmetry in the same way
as supersymmetry is broken in the imaginary-time formulation of thermal field theory.
The fermionic gluinos are subject to antiperiodic boundary conditions and thus
become massive at tree level, whereas adjoint scalars acquire masses through
loop corrections since they are not protected by gauge symmetry. In the limit
of large Kaluza-Klein mass scale, the only remaining degrees of freedom are the gauge bosons.
The dual geometry is given by a doubly Wick-rotated black hole
in AdS$_7\times S^4$,
\bea
ds^2&=&\frac{r^2}{L^2}\left( f(r)dx_4^2+\eta_{\mu\nu}dx^\mu dx^\nu
+dx_{11}^2 \right)\nonumber\\
&&+\frac{L^2}{r^2}\frac{dr^2}{f(r)}+\frac{L^2}{4}d\Omega_4^2,
\eea
with $f(r)=1-\R^6/r^6$ and a would-be thermal circle
\be
x^4\simeq x^4+2\pi R_4,\quad R_4\equiv\frac1\MKK=\frac{L^2}{3\R},
\ee
where the relation between $\R$ and $\MKK$ is determined by the absence
of a conical singularity at $r=\R$.
The background also has a Ramond-Ramond (R-R)
nonvanishing antisymmetric tensor gauge field with $N_c$
units of flux through the $S^4$.

The relation to the type IIA string-frame metric is
\be\label{ds210from11}
ds^2=G_{\hat M \hat N}dx^{\hat M} dx^{\hat N}=e^{-2\Phi/3}g_{MN}dx^M dx^N+e^{4\Phi/3}(dx^{11}+A_M dx^M)^2,
\ee
with $M,N=0,\ldots 9$ and $\hat M, \hat N$ additionally including the index 11. 
This leads to a nonconstant dilaton
$e^\Phi=(r/L)^{3/2}$ 
and $A_m=0$ for the above background geometry.

For later use we introduce the alternative radial coordinates 
$U\in(\UKK,\infty)$ and $Z\in(0,\infty)$, used also in Refs.~\cite{Sakai:2004cn,Sakai:2005yt}),
through
\be
U=\frac{r^2/2}{L},\quad
K(Z)\equiv 1+Z^2=\frac{r^6}{\R^6}=\frac{U^3}{\UKK^3}.
\ee
Note that the holographic boundary is at infinite values of $r$, $U$, and $Z$.

In terms of the radial coordinate $U$ the 10-dimensional metric reads
\be\label{ds210}
ds^2=\left(\frac{U}{\RSS}\right)^{3/2} \left[\eta_{\mu\nu}dx^\mu dx^\nu
+f(U)(dx^4)^2\right]+\left(\frac{\RSS}{U}\right)^{3/2}\left[\frac{dU^2}{f(U)}+U^2 d\Omega_4^2 \right]
\ee
with $f(U)=1-(\UKK/U)^3$; the nonconstant dilaton is given by 
\be\label{Phibackground}
e^\Phi=(U/\RSS)^{3/4}.
\ee

The parameters of the dual field theory 
are given by \cite{Kruczenski:2003uq,Sakai:2004cn,Sakai:2005yt,Kanitscheider:2008kd}
\footnote{This is based on a normalization of the Yang-Mills action as
$-\frac{1}{4 g_{\rm YM}^2}{\rm Tr}F_{\mu\nu}F^{\mu\nu}$,
which differs, however, from the convention used in particle physics,
where the coupling constant of SU($N_c$) gauge theories
is invariably defined as 
$\mathcal L=-\frac{1}{2 g^2}{\rm Tr}F_{\mu\nu}F^{\mu\nu}$
so that
$g^2=2 g_{\rm YM}^2$. This means
that the QCD coupling is given by $\alpha_s\equiv g^2/(4\pi)=
g_{\rm YM}^2/(2\pi)=\lambda/(2\pi N_c)$ in terms of
the 't Hooft coupling $\lambda\equiv N_c g^2$ as used here. 
Since we do not attempt to match with perturbative QCD here, this is of no
concern for the calculations performed below
(it is, however, important to take into account
when comparing quantitatively with weak-coupling results, see also
footnote 1 in Ref.~\cite{Blaizot:2006tk}).}
\bea\label{gYMNc}
\gYM^2=\frac{g_5^2}{2\pi R_4}=2\pi\gs\ls\MKK,\quad
(L/2)^3\equiv \RSS^3=\pi \gs \Nc \ls^3.\qquad
\eea
At scales much larger than $\MKK$, the dual theory turns into
5-dimensional super-Yang-Mills theory. However, it is not possible
to make $\MKK$ arbitrarily large without leaving the supergravity approximation.

The dual gauge theory exhibits confinement. Wilson loops connecting heavy quarks
at the boundary with large spatial separation along $x$ are represented by fundamental strings that
minimize their energy by having most of their length at minimal radial coordinate. The effective
string tension therefore tends to the value
\be\label{sigmastring}
\sigma=\frac1{2\pi l_s^2}\sqrt{-g_{tt}g_{xx}}\Big|_{U=\UKK}=
\frac1{2\pi l_s^2}\left(\frac{\UKK}{R}\right)^{3/2}=
\frac{2 g_{\rm YM}^2 N_c}{27\pi}\MKK^2.
\ee
In accordance with confinement, the dual theory has a mass gap for fluctuations of the background
geometry with scale set by $\MKK$.

\subsection{Holographic glueball spectrum}\label{secGB}

Ignoring all Kaluza-Klein modes on the compactification circles and all nontrivial harmonics on the $S^4$
with nonzero $R$ charge, the bosonic normal modes of the supergravity multiplet can be interpreted
as glueballs in the dual 3+1-dimensional Yang-Mills theory \cite{hep-th/9805129,Csaki:1998qr,
Hashimoto:1998if,Csaki:1999vb,Constable:1999gb,Brower:2000rp}.\footnote{In Ref.~\cite{Elander:2013jqa} this analysis was recently extended to modes obtained
by breaking the symmetry of the $S^4$.}
There are in total six independent wave equations for various scalar, vector, and tensor modes,
which were denoted as S$_4$, T$_4$, V$_4$, N$_4$, M$_4$, and L$_4$ in \cite{Brower:2000rp},
see Table \ref{tab1}.
These give three distinct possibilities to obtain modes with $J^{PC}=0^{++}$ quantum numbers,
corresponding to the 3+1-dimensional scalars $G_{11,11}$, $G_{4,4}$, and the $S^4$ volume fluctuation
$G^\alpha{}_\alpha$, where
the index $\alpha$ refers to the $S^4$. The latter, termed L$_4$ in Table \ref{tab1}, has
a lowest mass eigenvalue $\approx 3.57\MKK$ which is larger than those of all the other
wave equations and will be ignored in what follows.

The remaining two towers of scalar modes are described by the wave equations
denoted S$_4$ and T$_4$. The lowest mass eigenvalue is found in S$_4$, which corresponds asymptotically
to 11-dimensionally traceless metric fluctuations in
$G_{ii}$, $G_{11,11}$, and $G_{44}$.
The other scalar mode does not involve $G_{44}$ and can be attributed to the dilaton derived from $G_{11,11}$.
It is degenerate with the $2^{++}$ tensor mode (wave equation T$_4$) that is provided by transverse-traceless 
fluctuations in $G_{ij}$, $i,j=1,2,3$. (It is also degenerate with
the vector mode $1^{++}$ derived from $G_{11,i}$, but this mode is discarded as spurious from
the point of view of the 3+1-dimensional Yang-Mills theory because of negative ``$\tau$-parity'' \cite{Brower:2000rp},
implying that its dual operator is odd under a reflection $x^4\to-x^4$.)

Pseudoscalar ($0^{-+}$) modes are obtained from the 1-form field component $C_4$ 
descending from $G_{11,4}$ (wave equation V$_4$), whereas the 3-form field of 11-dimensional supergravity
is responsible for vector modes:
a vector $1^{+-}$ from the antisymmetric tensor field $B_{ij}$ (wave equation N$_4$), and a vector $1^{--}$ from the
3-form field components $C_{ij4}$ (wave equation M$_4$). All other modes can be
discarded due to negative $\tau$-parity.

\begin{figure}[t]
\centerline{\includegraphics[width=0.7\textwidth
]{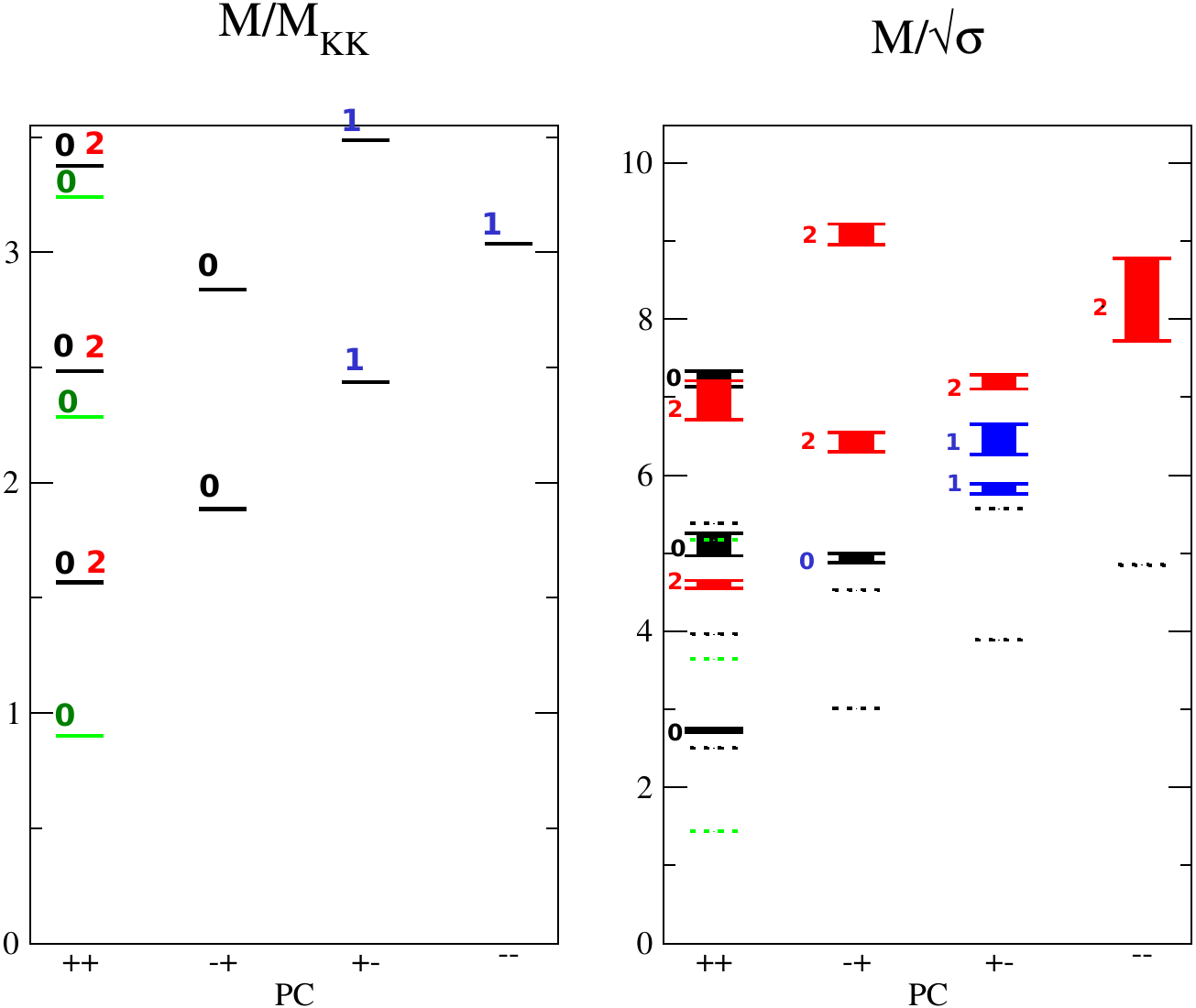}} 
\centerline{\small \hfil (a) \hfil\hfil (b) \hfil}
\medskip
\caption{
The glueball spectrum of the Witten model (a) in units of $\MKK$, (``exotic'' scalar
modes in green), compared to the spectrum obtained in the recent large-$N_c$ lattice
calculations of Ref.~\cite{Lucini:2010nv} (b) in
units of the square root of the string tension
$\sqrt{\sigma}$, juxtaposed such that the lowest tensor mode
is matched. The dotted lines in figure (b) give the glueball spectrum
of the Witten model when expressed in terms of the string tension
of the Witten model with the standard set of parameters (\ref{kappaSS})
for the Sakai-Sugimoto model.}
\label{figGB}       
\end{figure}

The glueball mass spectrum resulting from the numerical results listed in Table \ref{tab1} is displayed
in Fig.~\ref{figGB}, where it is compared with recent lattice results at large $N_c$ from Ref.~\cite{Lucini:2010nv},
which is in fact rather similar to that obtained for $N_c=3$ \cite{Morningstar:1999rf,Chen:2005mg}.
When juxtaposed such that the lowest tensor mode is matched, the holographic spectrum roughly reproduces the pattern
obtained in lattice gauge theory. Missing states of spin 2 with $PC\not=++$ and
higher spin states might be due to closed string modes.
On the other hand, there is a certain proliferation of $0^{++}$ states due to the existence of 
modes involving $G_{44}$, which have been termed ``exotic'' in Ref.~\cite{Constable:1999gb}, where
they were first considered. In fact, Ref.~\cite{Constable:1999gb} suspected that only one of the towers of scalar
states may survive in the limit $\MKK\to\infty$, where the Witten model would turn into an exact
string-gauge dual of large-$N_c$ Yang-Mills theory.

\begin{table}
\begin{tabular}{l|cccccc}
\toprule
Mode    &S$_4$&T$_4$&V$_4$&N$_4$&M$_4$&L$_4$\\
$J^{PC}$&$0^{++}$&$0^{++}/2^{++}$&$0^{-+}$&$1^{+-}$&$1^{--}$&$0^{++}$\\
\colrule
n=0&7.30835&22.0966&31.9853&53.3758&83.0449&115.002\\
n=1&46.9855&55.5833&72.4793&109.446&143.581&189.632\\
n=2&94.4816&102.452&126.144&177.231&217.397&277.283\\
n=3&154.963&162.699&193.133&257.959&304.531&378.099\\
n=4&228.709&236.328&273.482&351.895&405.011&492.171\\
\botrule
\end{tabular}
\caption{Our results for the mass spectrum $m_n^2$ of AdS$_7$
black hole metric fluctuations 
in the notation of \cite{Brower:2000rp}
(i.e.\ in units of $\R^2/L^4=\MKK^2/9$) obtained by
spectral methods cross-checked with a shooting method. The results for
the lowest modes agree completely with Ref.~\cite{Brower:2000rp}, while
for certain higher modes there are deviations in the last few digits.
$J^{PC}$ assignments are given only for the modes with even ``$\tau$-parity'' that
are expected to have a counterpart in QCD.}
\label{tab1}
\end{table}

\subsection{Normalization of glueball modes}

In order to be able to derive effective actions of the glueball modes and their interactions,
we need to calculate the normalization factors required for a canonical kinetic term.
For this purpose it is convenient to use the 11-dimensional
notation, where the fluctuations take their simplest form.

\subsubsection{Lowest (exotic) scalar glueball}

The lowest scalar glueball $0^{++}$ is associated with 
fluctuations involving asymptotically (for $r\to\infty$) 
$\delta G_{44}=-4 \delta G_{11}=-4 \delta G_{22}=-4 \delta G_{33}=-4 \delta G_{11,11}$.
In the bulk, other metric components are also involved, leading to the following
``exotic polarization'' \cite{Constable:1999gb}
\bea\label{deltaGG}
&&\delta G_{44} = -\frac{r^2}{L^2}f\HG(r)\G(x) \nonumber\\
&&\delta G_{\mu\nu} = \frac{r^2}{L^2}
\HG(r)\left[
\frac14 \eta_{\mu\nu} 
- \left(\frac14 + \frac{3\R^6}{5r^6-2\R^6}\right) 
\frac{\partial_\mu \partial_\nu}{\MG^2}
\right] \G(x),\nonumber\\
&&\delta G_{11,11} = \frac{r^2}{L^2}\frac14
 \HG(r)\G(x), \nonumber \\
&&\delta G_{rr} = -\frac{L^2}{r^2}f^{-1} \frac{3\R^6}{5r^6-2\R^6}
\HG(r) \G(x) ,\nonumber\\
&&\delta G_{r\mu} = \frac{90\, r^7 \R^6}{M_E^2 L^2 (5r^6-2\R^6)^2} 
\HG(r)\partial_\mu \G(x), 
\eea
where the eigenvalue equation is given by
\be
\frac1{r^3}\frac{d}{dr}r(r^6-\R^6)\frac{d}{dr} \HG(r)
+\left(\frac{432\,r^2\,\R^{12}}{(5r^6-2\R^6)^2}+L^4 \MG^2 \right) \HG(r)=0.
\ee

Integration over the $S^4$ reduces the 11-dimensional supergravity action to
\be
S=\frac1{2\kappa_{11}^2}(L/2)^4\Omega_4 \int d^7x \sqrt{-\det G}\left(R(G)+\frac{30}{L^2}\right)
\ee
with $2\kappa_{11}^2=(2\pi)^8 l_s^9 g_s^3$ and $\Omega_4=8\pi^2/3$. 

Inserting the metric fluctuations (\ref{deltaGG}) into the 7-dimensional action gives
\bea\label{deltaRG}
&&\int d^7x \sqrt{-\det G}\left.\left(R(G)+\frac{30}{L^2}\right)\right|_{\HG^2}
\nonumber\\
&=&
-\CG 
\int dx^{11}\,d^4x\,dx^4\, \frac12\left[(\6_\mu \G)^2+\MG^2 \G^2\right]
\eea
with
\be
\CG=\int_{\R}^\infty \frac{dr\,r^3}{L^3}\frac{5}{8}\HG(r)^2.
\ee
For the lowest eigenmode $\HG$ we obtain numerically
\be
\CG=0.057395\, [\HG(\R)]^2\frac{\R^4}{L^3}.
\ee
[This deviates from the result given in Ref.~\cite{Hashimoto:2007ze} by
a factor $\frac12$ that seems to be missing in their Eq.~(2.19).]

Requiring that upon integration over $x^4$ and $x^{11}$ the scalar field $\G(x)$
is canonically normalized leads to
\bea\label{Hnorm}
[\HG(\R)]^{-1}=[\HG(Z\!=\!0)]^{-1}&=&\frac1{\sqrt2}
0.0097839\,\lambda^{1/2}\,N_c\,\MKK \nonumber\\
&=&0.0069183\,\lambda^{1/2}\,N_c\,\MKK.
\eea
(This differs from \cite{Hashimoto:2007ze} only by the explicitly
written factor $1/{\sqrt2}$.) 


\subsubsection{Scalar and tensor modes from the tensor multiplet}

A scalar mode $0^{++}$ that does not involve metric components with index 4 is 
obtained from\footnote{As discussed recently in Ref.~\cite{Elander:2013jqa},
more possibilities for scalar (and other) glueball modes are obtained if
Ramond-Ramond field fluctuations which partially break the SO(5)
symmetry are included.}
\bea\label{deltaGD}
\delta G_{11,11}&=&-3\frac{r^2}{L^2}H_D(r)\D(x)\nonumber\\
\delta G_{\mu\nu}&=&\frac{r^2}{L^2}H_D(r)\left[\eta_{\mu\nu}-\frac{\6_\mu\6_\nu}{\Box}\right]\D(x).
\eea
Since upon reduction to 10 dimensions $\delta G_{11,11}$ is essentially the dilaton,
we shall refer to this mode as predominantly dilatonic. [Note that
also the ``exotic'' mode (\ref{deltaGG}) involves a dilaton component, but that
there the dominant component is $\delta G_{44}$. It should also be kept in mind
that the attribute ``exotic'' only refers to the holographic origin of this
mode, and not to any exotic $J^{PC}$ quantum numbers in the dual field theory.]

The tensor glueball $2^{++}$ is dual to metric fluctuations that have 
neither $\delta G_{44}$ nor $\delta G_{11,11}$, but
contain a transverse traceless polarization tensor in $\delta G_{\mu\nu}$. For example,
one can choose as only nonvanishing components
\be\label{deltaGT}
\delta G_{11}=-\delta G_{22}=-\frac{r^2}{L^2}H_T(r)G_T(x). 
\ee

The radial functions $H_{D,T}$ are determined by the equation
\be
\frac1{r^3}\frac{d}{dr}r(r^6-\R^6)\frac{d}{dr} H_{D,T}(r)
+L^4 M^2 H_{D,T}(r)=0,
\ee
with $M^2=M_D^2=M_T^2$.

Calculating the normalization of these glueball modes in analogy to (\ref{deltaRG}) leads to
\be
\mathcal C_{D,T}=\int_{\R}^\infty \frac{dr\,r^3}{L^3}\left\{ 6 H_D(r)^2 \atop H_T(r)^2 \right.
\ee
For the lowest eigenmode $\HG$ we obtain numerically
\be
\mathcal C_T=0.22547
\, [H_T(\R)]^2\frac{\R^4}{L^3}
\ee
and an analogous result for $\mathcal C_D$ with a coefficient 6 times as large.

This leads to
\bea\label{HDTnorm}
[H_{D,T}(\R)]^{-1}=[H_{D,T}(Z\!=\!0)]^{-1}=\lambda^{1/2}\,N_c\,\MKK 
\left\{ 0.033588 
\atop 0.013712 
\right.
\eea

\subsection{Glueball field/operator correspondence}

The above metric perturbations are sourced by operators in the dual field theory, which
is five-dimensional super-Yang-Mills theory compactified on the circle along $x^4$.

The operator dual to the tensor perturbations is simply the five-dimensional
energy-momentum tensor with three-dimensional indices. Omitting the adjoint scalars of the five-dimensional theory,
we have
\be
T_{mn}^{(5)}=T_{mn}^{\rm YM}+F_{4m}F_{4n}-\frac12 \delta_{mn} F_{4\mu}F_4{}^\mu+\ldots,
\ee
where $A_4$ is a further scalar that like the adjoint scalars of the five-dimensional
theory becomes massive through loop corrections.

The operators dual to the exotic and the predominantly dilatonic scalar modes can
be inferred from their couplings to the fields in the DBI action of D4 branes
in the limit of $r\to\infty$ \cite{Hashimoto:1998if}. The exotic scalar mode $\delta G^E_{MN}$
and the dilatonic one, $\frac14\delta G^D_{MN}$, turn out to source, respectively,\footnote{We
disagree here with Ref.~\cite{Brower:2000rp} which attributed $F_{\mu\nu}^2$ to $\delta G^D$
and $T_{00}$ to $\delta G^E$.}
\bea
\mathcal O^E&=&-\frac58 F_{\mu\nu}F^{\mu\nu}-\frac12 T_{00}^{\rm YM}+F_{4\mu}F_4{}^\mu-\frac12 F_{40}^2+\ldots,\\
\mathcal O^D&=&+\frac38 F_{\mu\nu}F^{\mu\nu}-\frac12 T_{00}^{\rm YM}+F_{4\mu}F_4{}^\mu-\frac12 F_{40}^2+\ldots.
\eea

The difference $\mathcal O^D-\mathcal O^E=F_{\mu\nu}F^{\mu\nu}$ is the purely four-dimensional glueball operator,
which is dual to $\frac14\delta G^D_{MN}-\delta G^E_{MN}$. However this linear combination is not a normal
mode in the gravitational background. We therefore need to keep the exotic and the predominantly dilatonic mode,
of which both, or perhaps only one of them, might correspond to the glueballs of the four-dimensional
Yang-Mills theory. To really end up with the latter, one would however need to
take the limit of large Kaluza-Klein mass $\MKK$, which is necessarily leaving the supergravity approximation.
In this limit, both modes will presumably receive important corrections. If one of the modes drops out of the spectrum,
one might suspect that it will more likely be $\delta G^E_{MN}$ as it includes a then spurious
polarization component $\delta G_{44}$.

In the following we shall consider both modes, as well as the tensor mode, when calculating
glueball-meson interactions within the Witten-Sakai-Sugimoto model, 
extending the analysis of Ref.~\cite{Hashimoto:2007ze}, which only
studied the lowest (exotic) $0^{++}$ mode.

\section{The Witten-Sakai-Sugimoto model}

Sakai and Sugimoto introduced chiral quarks in Witten's model
of pure-glue Yang-Mills theory by means of $N_f$
probe D8 and anti-D8 branes that fill all spatial directions except the
Kaluza-Klein circle \cite{Sakai:2004cn,Sakai:2005yt}. Quarks and antiquarks
are thus localized on separate points $x^4$ of the 4+1-dimensional boundary
theory. The global flavor symmetry $\mathrm U(N_f)_L\times \mathrm U(N_f)_R$
is however broken spontaneously,
because the subspace $x^4$-$U$ has the topology of a cigar
forcing the D8 and anti-D8 branes to join in the bulk.
The action of the joined D8 branes which describes the dynamics
of $q\bar q$ mesons through flavor gauge fields on the branes reads
\bea\label{SD8full}
S_{\rm D8}&=&-T_{\rm D8}\Tr\int d^9x e^{-\Phi}
\sqrt{-\det\left(\tilde g_{MN}+2\pi\alpha' F_{MN}\right)}+S_{\rm CS}\nonumber\\
&&=-(2\pi\alpha')^2T_{\rm D8}\Tr\int d^9x e^{-\Phi}\sqrt{-\tilde g}
\left(\mathbf 1+\frac14 \tilde g^{PR}\tilde g^{QS}F_{PQ}F_{RS}+O(F^4)\right)+S_{\rm CS}
\eea
with $T_{\rm D8}=(2\pi)^{-8}l_s^{-9}$, $\tilde g_{MN}$ the metric
on the 8+1-dimensional world volume induced by (\ref{ds210}), and $\Phi$ shifted such that
$e^\Phi=g_s (U/\RSS)^{3/4}$.
Because no backreaction of the D8 branes on the 10-dimensional background of the Witten model
is taken into account, this corresponds to the quenched approximation of QCD, as indeed
appropriate for the large-$N_c$ limit at fixed $N_f$. (For attempts to go beyond the
quenched approximation see Refs.~\cite{Burrington:2007qd,Bigazzi:2014qsa}.)

In the original version
of the Sakai-Sugimoto model that we shall use here, the D8 and anti-D8 branes are put
at antipodal points so that they join at the minimal value $U=\UKK$. In this case it is most
convenient to use the dimensionless coordinate $Z=\sqrt{(U/\UKK)^3-1}$ introduced already above,
but extended to the range $-\infty\ldots+\infty$ so that the radial integrations of the D8 and the anti-D8 branes are combined. The part of the DBI action quadratic in
the flavor field strength then reads
\be\label{SD8F2}
S_{\rm D8}^{(F^2)}=-\kappa\,\Tr\int d^4x \int_{-\infty}^\infty dZ\left[
\frac12 K^{-1/3}\eta^{\mu\rho}\eta^{\nu\sigma}F_{\mu\nu}F_{\rho\sigma}
+K\MKK^2\eta^{\mu\nu}F_{\mu Z}F_{\nu Z}\right]
\ee
with $K\equiv 1+Z^2$ and
\be
\kappa=(2\pi\alpha')^2 T_{\rm D8} g_s^{-1}\Omega_4 \frac13 \RSS^{9/2}\UKK^{1/2}
=\frac{\lambda N_c}{216\pi^3},
\ee
where (\ref{gYMNc}) as well as $\Omega_4=8\pi^2/3$ and $\MKK^2=(3/2)^2\UKK/\RSS^3$ have been used.

The Goldstone bosons of chiral symmetry breaking appear as
\be\label{SD80}
S_{\rm D8}=\frac{f_\pi^2}{4}\int d^4x\, \Tr\left( U^{-1}\6_\mu U \right)^2+\ldots,
\quad U=\mathrm P \exp\left\{ i\int_{-\infty}^\infty dZ A_Z\right\},
\ee
which determines the so-called pion decay constant in terms of $\lambda$ and $\MKK$ as
\be\label{fpi2}
f_\pi^2=\frac1{54\pi^4}\lambda N_c\MKK^2.
\ee

Massive vector and axial vector mesons arise as even and odd eigenmodes
of $A_\mu^{(n)}=\psi_n(Z) v^{(n)}_\mu(x)$ with eigenvalue equation
\be\label{psin}
-(1+Z^2)^{1/3}\6_Z\left( (1+Z^2)\6_Z \psi_n \right)=\lambda_n \psi_n,\quad \psi_n(\pm\infty)=0.
\ee
The lowest mode $v_\mu^{(1)}$ is 
interpreted as the isotriplet $\rho$ meson (or the $\omega$ meson
for the U(1) generator) with mass $m_\rho^2=\lambda_1 \MKK$
with the numerical result $\lambda_1=0.669314\ldots$.

The next-highest mode $v_\mu^{(2)}$ with eigenvalue $\lambda_2\approx 1.569$ 
is an axial vector that can be identified \cite{Sakai:2004cn} with the
meson $a_1(1260)$. The experimental value for the ratio $m_{a_1}/m_\rho\approx 1.59$
is remarkably close to $\sqrt(\lambda_2/\lambda_1)\approx 1.53$. 
Also the experimental value for the mass of the excited $\rho(1450)$
with $m_{\rho^*}/m_\rho\approx 1.89$ is 
is close to $\sqrt(\lambda_3/\lambda_1)\approx 2.07$. This
nice agreement may however be a bit fortuitous, since recent lattice simulations \cite{Bali:2013kia}
at large $N_c$, extrapolated to zero quark mass, give the higher values
$m_{a_1}/m_\rho\approx 1.86$ and $m_{\rho^*}/m_\rho\approx 2.40$.
This would correspond to errors 21\% and 16\%, respectively, which
may still be considered a success given that already the mass of $v_\mu^{(2)}$ is
above $\MKK$. (For more checks of the quantitative predictions of
the Witten-Sakai-Sugimoto model see Ref.~\cite{Rebhan:2014rxa}.)
Optimistically, one can therefore hope that the Witten-Sakai-Sugimoto
model is a useful approximation to QCD up to masses of two or three times $\MKK$.

\subsection{Choice of parameters}

Matching the result for the $\rho$ meson mass with its experimental value, $m_\rho=\sqrt{\lambda_1}\MKK\approx 776$~MeV,\footnote{The mass of the
$\omega$ meson, which is degenerate with the $\rho$ meson in the Sakai-Sugimoto
model, is only slightly higher in real QCD.}  
fixes the Kaluza-Klein mass to \cite{Sakai:2004cn,Sakai:2005yt} $\MKK=949$~MeV.
This determines the masses of the other vector and axial vector mesons, which come out
in rough agreement with experiment.
The masses of the lowest (exotic) and the predominantly dilatonic 
scalar glueball, the tensor glueball (degenerate
with the dilatonic scalar), 
and the lowest pseudoscalar glueball are fixed to,
respectively,
\bea\label{MGs}
&&\MG=\sqrt{7.30834/9}\,\MKK\approx 855\,{\rm MeV},\nonumber\\
&&M_D=M_T=\sqrt{22.0966/9}\,\MKK\approx 1487\,{\rm MeV},\nonumber\\
&&M_P=\sqrt{31.9853/9}\,\MKK\approx 1789\,{\rm MeV},\nonumber\\
&&M_{E^*}=\sqrt{46.9855/9}\,\MKK\approx 2168\,{\rm MeV},\nonumber\\
&&M_{D^*}=M_{T^*}=\sqrt{55.5833/9}\,\MKK\approx 2358\,{\rm MeV},
\eea
where we have also given the masses of some of the corresponding excited states (marked by a star).

The lowest scalar glueball involving the exotic polarization (\ref{deltaGG}) with a dominant $\delta G_{44}$
component is found to be only 10\% heavier than the $\rho$ meson. This is in stark contrast to lattice results both
for quenched $N_c=3$ and $N_c=\infty$ QCD \cite{Lucini:2010nv}, where the lightest glueball is about twice as heavy.

A possible modification of the Sakai-Sugimoto model consists of
choosing a nonmaximal separation of the D8-$\overline{\rm D8}$ branes \cite{Antonyan:2006vw,Aharony:2006da}.
The latter then join at a value $U=U_0>\UKK$ and the mass of a string stretched
between $\UKK$ and $U_0$ has been interpreted as a ``constituent'' quark mass.
Unfortunately, this only makes the problem worse: Nonmaximal separation
increases the eigenvalue $\lambda_1$ \cite{Peeters:2007ab} while the glueball spectrum
is unaffected. With a constituent quark mass of 310~MeV and keeping
the mass of the $\rho$ meson fixed as done in Ref.~\cite{Callebaut:2011ab}, $\MKK$ is reduced to
720~MeV, which reduces all values in (\ref{MGs}) by 25\%.

With maximal separation and the standard choice $\MKK=949$~MeV, the mass of the dilatonic glueball is
not far from the numerical result obtained in lattice gauge theory for the lightest scalar glueball state,
while a degeneracy with the tensor glueball is not observed there---the latter is instead significantly heavier.
This degeneracy might perhaps be lifted by higher-derivative corrections when going beyond the leading
supergravity approximation. Similarly, it is conceivable that only the dilatonic glueball survives
in the (unfortunately inaccessible) limit to a complete holographic QCD and that therefore the lowest
scalar mode is to be discarded. We shall come back to this question when calculating the decay width
of the various glueball states.

In order to calculate glueball-meson interactions, we shall need to extrapolate
to finite coupling and finite $N_c=3$.
The original \cite{Sakai:2004cn,Sakai:2005yt} and most widely used choice is
obtained from matching $f_\pi\approx 92.4\,{\rm MeV}$ in (\ref{fpi2}) which gives
\be\label{kappaSS}
\kappa\equiv
\lambda N_c/(216\pi^3)=7.45\cdot10^{-3}
\;\Rightarrow\; \lambda\approx 16.63 \quad (N_c=3).
\ee
[The original and published version of Ref.~\cite{Sakai:2004cn,Sakai:2005yt}
contained an error in the prefactor of the D8 brane action for $N_f>1$
involving a different definition of $\kappa$, which led
to a 't Hooft coupling of about 8.3 and effectively
a correspondingly reduced pion decay constant.
This error,
which was later corrected in the e-print versions of Ref.~\cite{Sakai:2004cn,Sakai:2005yt}, did not
affect the mass spectra of mesons
obtained in Ref.~\cite{Sakai:2004cn,Sakai:2005yt}, but
it does affect all interactions. Unfortunately,
Ref.~\cite{Hashimoto:2007ze} still employed the incorrectly
matched 't Hooft coupling, affecting
all meson and glueball decay rates calculated therein.]

In what follows, we shall take (\ref{kappaSS}) as the standard choice, but also
consider as an alternative a value of the 't Hooft coupling obtained by
matching $m_\rho/\sqrt{\sigma}$, where $\sigma$ is the string tension (\ref{sigmastring}), to
the large-$N_c$ lattice result of Ref.~\cite{Bali:2013kia}.
Ref.~\cite{Bali:2013kia} obtained $m_\rho/\sqrt{\sigma}=1.504(50)$, whose central value corresponds to
$\lambda=12.55$. With the ``standard'' value $\lambda\approx 16.63$ the Sakai-Sugimoto model
predicts $m_\rho/\sqrt{\sigma}\approx 1.306$, which agrees within 15\% but points to a smaller
't Hooft coupling and thus a smaller string tension. A smaller 't Hooft coupling has also been
argued for in Ref.~\cite{Imoto:2010ef}, where the spectrum of higher-spin mesons obtained
from massive open string modes has been considered. We shall therefore consider a downward
variation of $\lambda\approx 16.63\ldots 12.55$ to get an idea of the variability of the predictions
of the Witten-Sakai-Sugimoto model.

Before turning to decay rates, we consider two other predictions of the Witten-Sakai-Sugimoto model
at finite $N_c$ where the concrete value of $\lambda$ matters.

At infinite $N_c$, the Goldstone bosons include also a massless $\eta'$ pseudoscalar meson
from the spontaneous breaking of the $\mathrm U_A(1)$ symmetry, whose anomaly is suppressed
at $N_c\to\infty$. However, at finite $N_c$, the Sakai-Sugimoto model predicts
a finite mass for the $\eta'$ meson through a Witten-Veneziano formula evaluated
already in \cite{Sakai:2004cn} with the result
\be\label{metaprime}
m_{\eta'}=\frac1{3\sqrt3 \pi}\sqrt{\frac{N_f}{N_c}}\lambda\MKK.
\ee
With $\MKK=949$~MeV and $\lambda\approx 16.63$ (or 12.55) the numerical value
for $N_c=N_f=3$ turns out to be 967~MeV (730~MeV). The higher value is surprisingly
close to the experimental value 958~MeV, but actually a smaller value than that might
perhaps be expected given the absence of a strange quark mass. At any rate,
the right ballpark seems to be reached with the parameters considered here.

Another quantity of interest, in particular in connection with glueball physics, is
the gluon condensate which was calculated in Ref.~\cite{Kanitscheider:2008kd} as
\be\label{gluoncondensate}
C^4\equiv\left<\frac{\alpha_s}{\pi}G_{\mu\nu}^a
G^{a\mu\nu}\right>=\frac{4N_c}{3^7\pi^4}\lambda^2\MKK^4.
\ee
For $\lambda\approx 16.63$ this yields $C^4=0.0126\,{\rm GeV}^4$, almost
identical to the standard SVZ sum rule value \cite{Shifman:1978bx}, while
for $\lambda=12.55$ a significantly smaller value of 0.0072 GeV$^4$ is obtained.
Using sum rules both smaller \cite{Ioffe:2005ym} and larger \cite{Narison:2011xe} values than the standard SVZ sum rule value 
are discussed in the literature, 
while lattice simulations typically give significantly larger values, 
which are however of the same size as ambiguities from the subtraction procedure \cite{Bali:2014sja}.
While a quantitative comparison thus does not seem to be in order, we note that
the gluon condensate is predicted to be small. 

\subsection{Normalization of $q\bar q$ modes}

For the calculation of decay rates we will initially consider $N_f=2$, dropping
the strange quark whose nonnegligible mass cannot be easily accommodated
within the Sakai-Sugimoto model (see however Ref.~\cite{0708.2839,Aharony:2008an,Hashimoto:2008sr,McNees:2008km});
the possible effects of the finite quark masses will be discussed in Section \ref{sec:extrapol}.

In the chiral Sakai-Sugimoto model, the Goldstone bosons are the massless pions
contained in 
\be
A_Z=\UKK\phi_0(Z)\pi(x^\mu), 
\ee
where $\UKK$ has been included to
render the mode function $\phi_0(Z)$ dimensionless.\footnote{For our purposes it is most convenient
to keep $A_Z$ nonzero. The frequently adopted gauge choice $A_Z=0$ leads to a different
but physically equivalent
field parametrization of the Goldstone bosons.}
The U(1) part of $A_Z$ corresponds to the $\eta'$ meson, which is a Goldstone boson only at infinite $N_c$;
for finite $N_c$ it receives a mass through the Witten-Veneziano mechanism \cite{Sakai:2004cn} (see Eq.~(\ref{metaprime}) below).

The only vector mesons that we shall consider will be
the isotriplet $\rho$ meson described by the traceless part of
\be
A_\mu=\psi_1(Z)\rho_\mu(x^\nu),
\ee
and the isosinglet $\omega$ meson given by the corresponding expression proportional to the unit matrix.

Following Ref.~\cite{Hashimoto:2007ze} (which here differs from \cite{Sakai:2004cn,Sakai:2005yt})
we choose the generators of the SU(2) flavor group such that $\Tr T^a T^b=\delta^{ab}$. Canonical
normalization of the fields $\pi^a$ and $\rho^a_\mu$ in (\ref{SD8F2})
such that upon integration over $Z$ one has
\be
S=-\Tr\int d^4x\left[\frac12(\6_\mu\pi)^2+\frac14 F_{\mu\nu}^2+
\frac12 \lambda_1 \MKK^2\rho_\mu^2+\ldots\right]
\ee
leads to
\bea
&&2\kappa\int_{-\infty}^\infty dZ\,K^{-1/3}(\psi_1)^2=1, \\
&&2\kappa(\UKK\MKK)^2 \int_{-\infty}^\infty dZ\,K(\phi_0)^2=1.
\eea
The first relation determines the value of $\psi_1$ at $Z=0$
with the help of the numerical result
\be
\int_{-\infty}^\infty dZ\,K^{-1/3}(\psi_1)^2=2.80302\ldots \, \psi_1^2(0),
\ee
while the second fixes the normalization of $\phi_0\propto 1/K\equiv 1/(1+Z^2)$ as
\be
\UKK\MKK\,\phi_0=\frac{1}{\sqrt{2\pi\kappa}} \frac1K.
\ee

\subsection{$\rho$ and $\omega$ meson decay}


The $\rho$-$\pi$ interactions 
are determined by the second term of (\ref{SD8F2}), using\footnote{Here we
follow the conventions of Ref.~\cite{Hashimoto:2007ze}. Note that in Ref.~\cite{Sakai:2004cn,Sakai:2005yt} the matrix-valued flavor gauge fields are
antihermitean.} 
$F_{\mu Z}=\6_\mu A_Z - \6_Z A_\mu-i[A_\mu,A_Z]$. The effective vertex between the four-dimensional fields $\rho$ and $\pi$'s are
obtained upon integration of the resulting products of the mode functions $\psi_1(Z)$ and $\phi_0\propto 1/K$.
For the process $\rho\to\pi\pi$ we need specifically
\be 
\mathcal L_{\rho\pi\pi}=-g_{\rho\pi\pi} 
\epsilon_{abc}(\partial_\mu \pi^a)
\rho^{b\mu}\pi^c,\quad
g_{\rho\pi\pi}=\sqrt2
\int dZ \frac1{\pi K}\psi_1=\sqrt{2}\times 24.030
\,\lambda^{-\frac12} N_c^{-\frac12}.
\ee
This agrees with the numerical value given in table 3$\cdot$34 of 
Ref.~\cite{Sakai:2005yt}
for $g_{v^1\pi\pi}\equiv g_{\rho\pi\pi}$. 
($g_{\rho\pi\pi}/\sqrt2$ 
was denoted as $c_6$ in \cite{Hashimoto:2007ze}; we will
reserve $c_i$, $i=1,2,3\ldots$, for the coefficients in the interactions of
the glueball field with mesons, for which we will follow the
conventions chosen in \cite{Hashimoto:2007ze}.)

The amplitude for the decay of a $\rho$ meson at rest
with polarization $\epsilon^\mu=(0,\mathbf e)$
into two pions with momenta 
$p^\mu=(|\mathbf p|,\mathbf p)$ and
$q^\mu=(|\mathbf p|,-\mathbf p)$
reads
\be
\mathcal M=ig_{\rho\pi\pi}\, \epsilon^\mu(p_\mu-q_\mu)
=2ig_{\rho\pi\pi}\, \mathbf e\cdot \mathbf p.
\ee
The expression for the decay rate involves a directional average, leading to
\be\label{Gammarho}
\Gamma_\rho/m_\rho
=\frac1{4\pi}\int d\Omega \frac{|\mathcal M|^2}{16\pi m_\rho^2}=
\frac{g_{\rho\pi\pi}^2}{48\pi}\approx \frac{7.659}{\lambda N_c}
\approx \left\{ 0.1535 \; (\lambda=16.63) \atop 0.2034 \; (\lambda=12.55) \right.
\ee
which compares remarkably well with the current experimental
value $\Gamma_\rho/m_\rho= 0.191(1)$ from Ref.~\PDG\ (although it should
be noted that in this process the finite pion mass implies a reduction
by about 20\% compared to a decay into massless particles so that the coupling
$g_{\rho\pi\pi}$ appears somewhat underestimated with our
range of parameters for the Sakai-Sugimoto model).

The decay of the $\omega$ meson into $\pi^0\gamma$ and $\pi^0\pi^+\pi^-$, which
is due to the Chern-Simons part of the D8 brane action, has been calculated
in \cite{Sakai:2005yt}, with the result $2.58$~MeV for the dominant 3-pion decay,
which is significantly below the experimental value $\approx 7.6$~MeV.
However, the result of \cite{Sakai:2005yt} is proportional to $\lambda^{-4}$.
Varying again $\lambda$ from 16.63 to 12.55 gives the range 2.58\ldots7.96~MeV, which
happens to include the experimental value.

So the model appears to make reasonable semi-quantitative estimates for meson interactions,
which is quite remarkable given that after fixing the mass scale and setting $N_c=3$,
there is only one free parameter, namely $\lambda$. This certainly
makes it interesting to consider the predictions of this model for glueball decay rates in detail.

\section{Glueball-meson interactions}

The glueball modes, which have been obtained in Sect.~\ref{secGB} in terms of
11-dimensional metric perturbations $\delta G_{\hat M \hat N}$, translate to perturbations of the type-IIA
string metric $g_{MN}$ and the dilaton $\Phi$
according to (\ref{ds210from11}). Explicitly, this gives
\bea\label{deltag10}
g_{\mu\nu}&=& \frac{r^3}{L^3}\left[ \left(1+\frac{L^2}{2r^2} \delta G_{11,11}\right)\eta_{\mu\nu} +\frac{L^2}{r^2} \delta G_{\mu\nu} \right],\nonumber\\
g_{44}&=& \frac{r^3f}{L^3}\left(1+\frac{L^2}{2r^2} \delta G_{11,11}
+\frac{L^2}{r^2 f}\delta G_{44}\right),\nonumber\\
g_{rr}&=& \frac{L}{rf}\left(1+\frac{L^2}{2r^2} \delta G_{11,11} + \frac{r^2 f}{L^2} \delta G_{rr} \right),\nonumber\\
g_{r\mu}&=& \frac{r}{L} \delta G_{r\mu} ,\nonumber\\
g_{\Omega\Omega}&=&\frac{r}{L} \left(\frac{L}{2}\right)^2 
\left(1+\frac{L^2}{2r^2} \delta G_{11,11}\right) ,\nonumber\\
e^{4\Phi/3}&=&\frac{r^2}{L^2}\left(1+\frac{L^2}{r^2}\delta G_{11,11}\right).
\eea
Here we differ from Ref.~\cite{Hashimoto:2007ze} where the metric
fluctuations $g_{\Omega\Omega}$ on the $S^4$ have been omitted. 
As one can check (Appendix \ref{sec:tendimfieldeq}), the 10-dimensional 
equations for the glueball modes
are satisfied only when the fluctuation in $g_{\Omega\Omega}$ is kept.\footnote{%
In 10 dimensions, the induced fluctuations in $g_{\Omega\Omega}$ are
in fact necessary to decouple the mode L$_4$, which
in 11 dimensions corresponds
to pure $S^4$ volume fluctuations, as can be seen from
the explicit 10-dimensional calculations in Ref.~\cite{Hashimoto:1998if}.}

We shall consider in turn the lowest glueball dual to the metric fluctuations (\ref{deltaGG}), referred to as
``exotic'' because it involves $\delta G_{44}$ besides dilaton fluctuations in $\delta G_{11,11}$, 
the predominantly dilatonic
glueball associated to (\ref{deltaGD}), and the tensor glueball with metric fluctuations (\ref{deltaGT}).
Inserting the respective metric fluctuations in the D8 brane action and integrating over the bulk coordinates
yields effective interaction Lagrangians which are given in full detail in Appendix \ref{sec:Lglueint}. 

\subsection{Glueball decay to two pions}\label{sec:G2pi}

\begin{figure}[t]
\includegraphics[width=0.3\textwidth]{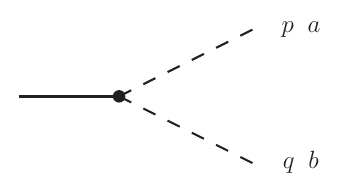}
\caption{Leading-order glueball decay into two pions.}
\label{fig:g2pi}
\end{figure}

The effective, 3+1-dimensional interaction Lagrangian for the lowest (exotic) $0^{++}$ glueball
$\G$ reads (omitting terms that vanish when $\G$ is on-shell)
\be
\mathcal L^{\G\to\pi\pi}=-\Tr\left[\frac12 c_1 \6_\mu\pi \6_\nu\pi
\frac{\6^\mu\6^\nu}{\MG^2}\G
+\frac12 \breve c_1 \6_\mu\pi \6^\mu\pi\, \G\right]
\ee
with coupling constants $c_1$ and $\breve c_1$ defined in
(\ref{cbrevec}) and numerically given in Table \ref{tabcG}.

\begin{table}
\begin{tabular}{l|c|r}
\toprule
&vertex&value\\
\colrule
$c_1/\sqrt{2}$ & $\G\partial\pi\partial\pi$ & $44.304 
\; \lambda^{-\frac12}\, N_c^{-1} \MKK^{-1}$ \\
$c_2/\sqrt{2}$ & $\G\rho\rho$ & $5.0318\; \lambda^{-\frac12}\, N_c^{-1} \MKK^{-1}$ \\
$c_3/\sqrt{2}$ & $\G\partial\rho\partial\rho$ & $49.334\; \lambda^{-\frac12}\, N_c^{-1} \MKK^{-1}$ \\
$c_4/\sqrt{2}$ & $\G\rho\partial\rho$ & $-7.4810\; \lambda^{-\frac12}\, N_c^{-1} \MKK^{+1}$ \\
$c_5/\sqrt{2}$ & $\G\rho\pi\partial\pi$ & $1428.1 \; \lambda^{-1}\, N_c^{-\frac32} \MKK^{-1}$ \\
$\breve c_1/\sqrt{2}$ & $\G\partial\pi\partial\pi$ & $11.590 
\; \lambda^{-\frac12}\, N_c^{-1} \MKK^{-1}$ \\
$\breve c_2/\sqrt{2}$ & $\G\rho\rho$ & $2.0970\; \lambda^{-\frac12}\, N_c^{-1} \MKK^{-1}$ \\
$\breve c_3/\sqrt{2}$ & $\G\partial\rho\partial\rho$ & $12.814\; \lambda^{-\frac12}\, N_c^{-1} \MKK^{-1}$ \\
$\breve c_5/\sqrt{2}$ & $\G\rho\pi\partial\pi$ & $359.33 \; \lambda^{-1}\, N_c^{-\frac32} \MKK^{-1}$ \\
\colrule
$c_1^*/\sqrt{2}$ & $\G^*\partial\pi\partial\pi$ & $24.641
\; \lambda^{-\frac12}\, N_c^{-1} \MKK^{-1}$ \\
$c^*_2/\sqrt{2}$ & $\G^*\rho\rho$ & $-0.8227
\; \lambda^{-\frac12}\, N_c^{-1} \MKK^{-1}$ \\
$c^*_3/\sqrt{2}$ & $\G^*\partial\rho\partial\rho$ & $27.906
\; \lambda^{-\frac12}\, N_c^{-1} \MKK^{-1}$ \\
$c^*_4/\sqrt{2}$ & $\G^*\rho\partial\rho$ & $-1.7467
\; \lambda^{-\frac12}\, N_c^{-1} \MKK^{+1}$ \\
$c^*_5/\sqrt{2}$ & $\G^*\rho\pi\partial\pi$ & $858.66
\; \lambda^{-1}\, N_c^{-\frac32} \MKK^{-1}$ \\
$\breve c^*_1/\sqrt{2}$ & $\G^*\partial\pi\partial\pi$ & $4.5843
\; \lambda^{-\frac12}\, N_c^{-1} \MKK^{-1}$ \\
$\breve c^*_2/\sqrt{2}$ & $\G^*\rho\rho$ & $-1.2390
\; \lambda^{-\frac12}\, N_c^{-1} \MKK^{-1}$ \\
$\breve c^*_3/\sqrt{2}$ & $\G^*\partial\rho\partial\rho$ & $5.3829
\; \lambda^{-\frac12}\, N_c^{-1} \MKK^{-1}$ \\
$\breve c^*_5/\sqrt{2}$ & $\G^*\rho\pi\partial\pi$ & $176.99
\; \lambda^{-1}\, N_c^{-\frac32} \MKK^{-1}$ \\
\botrule
\end{tabular}
\caption{Coupling coefficients in interaction Lagrangian of lowest glueball.
(Here we give numerical results for $c_i/\sqrt{2}$
to permit a comparison with the results listed in \cite{Hashimoto:2007ze},
with which we disagree by a factor of $\sqrt{2}$ in (\ref{Hnorm}). 
Taking this into account we agree with all
numerical values, with the exception of
$c_4$, which in \cite{Hashimoto:2007ze} seems to be missing 
the numerical factor contained in the normalization of $\HG(Z)$.)
The coefficients $\breve c_{1,2,3,5}$ are coupling constants due to
$S_4$ volume fluctuations induced by the lowest glueball that were
apparently dropped in \protect\cite{Hashimoto:2007ze}. 
Coefficients with a star
indicate the corresponding constants for the first excited exotic mode.}
\label{tabcG}
\end{table}

\begin{table}
\begin{tabular}{l|c|r}
\toprule
&vertex&value\\
\colrule
$d_1$ & $\tilde G\partial\pi\partial\pi$ & $17.226
\; \lambda^{-\frac12}\, N_c^{-1} \MKK^{-1}$ \\
$d_2$ & $\tilde G\rho\rho$ & $4.3714 \; \lambda^{-\frac12}\, N_c^{-1} \MKK^{-1}$ \\
$d_3$ & $\tilde G\partial\rho\partial\rho$ & $18.873\; \lambda^{-\frac12}\, N_c^{-1} \MKK^{-1}$ \\
$d_5$ & $\tilde G\rho\pi\partial\pi$ & $512.20 \; \lambda^{-1}\, N_c^{-\frac32} \MKK^{-1}$ \\
\colrule
$d^*_1$ & $\tilde G^*\partial\pi\partial\pi$ & $11.906 
\; \lambda^{-\frac12}\, N_c^{-1} \MKK^{-1}$ \\
$d^*_2$ & $\tilde G^*\rho\rho$ & $-0.9415\; \lambda^{-\frac12}\, N_c^{-1} \MKK^{-1}$ \\
$d^*_3$ & $\tilde G^*\partial\rho\partial\rho$ & $13.680 \; \lambda^{-\frac12}\, N_c^{-1} \MKK^{-1}$ \\
$d^*_5$ & $\tilde G^*\rho\pi\partial\pi$ & $419.46 \; \lambda^{-1}\, N_c^{-\frac32} \MKK^{-1}$ \\
\botrule
\end{tabular}
\caption{Coupling coefficients $d_i$ ($t_i\equiv\sqrt6\,d_i$)
in the interaction Lagrangian of the lowest glueballs in the tensor multiplet (dilaton and tensor), 
collectively denoted as $\tilde G$, with a star indicating the first excited mode.
(Note that there is no term
analogous to the one involving $c_4$ for the lowest (exotic) glueball.)
}
\label{tabct}
\end{table}

The corresponding result for the dilatonic scalar $0^{++}$ and the $2^{++}$ mode,
denoted $\D$ and $T^{\mu\nu}$, respectively, is
\bea
\mathcal L^{\D\to\pi\pi}&=& \frac{1}{2}d_1 \Tr \partial_\mu\pi\partial_\nu\pi 
\left(\eta^{\mu\nu}-\frac{\partial^\mu \partial^\nu}{M_D^2}\right)\D,\\
\mathcal L^{G_T\to\pi\pi}&=& \frac12 t_1  
\Tr\6_\mu\pi\6_\nu\pi \, T^{\mu\nu},
\quad t_1\equiv \sqrt6\,d_1.
\eea
$\D$ is a canonically normalized real scalar,
and $T^{\mu\nu}$ a massive tensor field with transverse traceless
polarizations, normalized such that 
\be
\mathcal L^T=\frac14 T_{\mu\nu}(\Box-M_T^2)
T^{\mu\nu}+B_\mu \6_\nu T^{\mu\nu}+B \eta_{\mu\nu} T^{\mu\nu} +\ldots,
\ee
where $B_\mu$ and $B$ are Lagrange multiplier fields.
The coefficient $d_1$ is given
in Table \ref{tabct}.


For the two scalar glueballs described by $\G$ and $\D$, the
decay width into two pions is given by the simple expression
\be
\Gamma_{G_{E,D}\to\pi\pi}=\frac{|\mathbf p|}{8\pi M^2_{E,D}}|\mathcal M_{E,D}|^2\times3\times\frac12,
\ee
where $\mathbf p$ is the momentum of one of the pions
in the rest frame of the glueball with $|\mathbf p|=M_{E,D}/2$, the factor of 3 comes from the sum
over the isospin quantum number, 
and the factor of $\frac12$ is included because the two pions
are identical. The amplitude for the decay of $\G$ and $\D$
is, respectively,
\bea
|\mathcal M_E|&=&|(c_1+\breve c_1) p_0 q_0-\breve c_1 \mathbf p\cdot\mathbf q|=|c_1+2\breve c_1|\frac{M^2_E}{4},\\
\quad |\mathcal M_D|&=&|d_1 \mathbf p\cdot\mathbf q|=
|d_1|\frac{M^2_D}{4}.
\eea

For the tensor glueball an average over the polarizations
of the tensor is needed. Alternatively, we can choose a
fixed polarization $\epsilon^{11}=-\epsilon^{22}=1$ and
integrate over the orientation of our Cartesian coordinates.
This leads to the scattering amplitude (in the rest frame
of the tensor glueball)
\be\label{MTpipi}
|\mathcal M_T|=|t_1(p_x^2-p_y^2)|,\quad |\mathbf p|=M_T/2,
\ee
and the decay width
\be\label{GTpipi}
\Gamma_{T\to\pi\pi}=\frac{|\mathbf p|}{8\pi M_T^2}\int \frac{d\Omega}{4\pi}|\mathcal M_T|^2
\times\frac32=\frac1{640\pi}
|t_1|^2 M_T^3.
\ee

Numerically we obtain\footnote{Ignoring the contribution
involving $\breve c_1$, the result for
the relative width of the scalar glueball $G$ would read $0.040$ in agreement
with the result of \cite{Hashimoto:2007ze}, because
the fact that the coefficient in $|c_1|^2$ is twice that of
\cite{Hashimoto:2007ze} is exactly compensated
by $\lambda^{-1}$ in $|c_1|^2$ being half that in \cite{Hashimoto:2007ze}.}
with $\lambda\approx 16.63$
for the scalar glueballs $\G$, $\D$, and $\Ds$ 
\bea
\label{GEtopipi}
\Gamma_{\G\to\pi\pi}/\MG&=&\frac{3|c_1+2\breve c_1|^2 \MG^2}{512\pi}\approx
\frac{13.79}{\lambda N_c^2}\approx 0.092
\quad(\MG\approx 855{\rm MeV})\\
\Gamma_{\D\to\pi\pi}/M_D&=&\frac{3|d_1|^2 M_D^2}{512\pi}\approx 
\frac{1.359}{\lambda N_c^2}\approx 0.009
\quad(M_D\approx 1487{\rm MeV})\\
\label{Ds2piwidth}
\Gamma_{\Ds\to\pi\pi}/M_{D^*}&=&\frac{3|d^*_1|^2 M_{D^*}^2}{512\pi}\approx
\frac{1.633}{\lambda N_c^2}\approx 0.011
\quad(M_{D^*}\approx 2358{\rm MeV})\eea
and for the tensor
\bea\label{Ttopipi}
\Gamma_{T\to\pi\pi}/M_T&=&\frac{|t_1|^2 M_T^2}{640\pi}\approx
\frac{2.174}{\lambda N_c^2}\approx0.0145
\quad(M_T\approx 1487{\rm MeV})\\
\label{Tstopipi}
\Gamma_{T^*\to\pi\pi}/M_{T^*}&=&\frac{|t_1^*|^2 M_{T^*}^2}{640\pi}\approx
\frac{2.613}{\lambda N_c^2}\approx0.0175
\quad(M_{T^*}\approx 2358{\rm MeV})
\eea

If we replace the standard choice $\lambda\approx 16.63$ by the smaller
value 12.55 as discussed above, all these decay rates which are proportional to $\lambda^{-1}$
increase by 33\% (see Table \ref{tab2pions} for a summary).

\begin{table}
\begin{tabular}{l|r|c}
\toprule
& $M$ & $\Gamma/M$ \\ 
\colrule
$\G\to2\pi$  & 855 & 0.092 \ldots 0.122 \\
$\G^*\to2\pi$  & 2168 & 0.149 \ldots 0.197 \\ 
\colrule
$\D\to2\pi$ & 1487 & 0.009 \ldots  0.012 \\ 
$\Ds\to2\pi$ & 2358 & 0.011 \ldots 0.014 \\ 
\colrule
$T\to2\pi$ & 1487 & 0.0145  \ldots 0.0193 \\ 
$T^*\to2\pi$ & 2358 & 0.0175 \ldots  0.0233 \\ 
\botrule
\end{tabular}
\caption{Decay width of scalar and tensor glueballs into 2 (massless) pions divided by glueball mass
for $\lambda=16.63\ldots12.55$}
\label{tab2pions}
\end{table}

A somewhat anomalous feature of the lowest (exotic) scalar glueball is that its
width is much larger than the next-to-lowest (dilatonic) scalar glueball while
having a rather low mass. 
This appears rather unnatural if the dilatonic scalar glueball is interpreted as an
excited scalar glueball and
may be another indication that the exotic mode should be discarded altogether.

Interestingly enough, a scenario with a broad glueball around 1 GeV in combination
with a narrow glueball in the range predicted by quenched (as well as unquenched \cite{Gregory:2012hu})
lattice gauge theory has been
proposed in Ref.~\cite{Narison:1996fm,Narison:2005wc,Mennessier:2008kk,Kaminski:2009qg} on the basis of
QCD spectral sum rules. There the lighter glueball, called $\sigma_{\rm B}$, plays the role of an important
bare glueball component of the $\sigma$-meson $f_0(500)$, while a higher narrow glueball
around 1.5-1.6 GeV is required by the consistency of subtracted and unsubtracted sum rules.
The glueball state $\sigma_{\rm B}$ of Ref.~\cite{Narison:1996fm,Narison:2005wc,Kaminski:2009qg}
has a broad decay width into two pions, in fact even much broader than (\ref{GEtopipi}), which
makes us speculate that the exotic scalar glueball of the Witten-Sakai-Sugimoto model
could find a role as the holographic dual of a pure-glue component of the $\sigma$-meson, perhaps while having
to be discarded from the spectrum of the pure pure-glue Witten model.\footnote{This dichotomy might be
due to the fact that the flavor D8 branes of the Sakai-Sugimoto model are localized in the $x^4$
direction along which the graviton mode associated with $\G$ is polarized,
whereas this extra spatial direction should play no active role in the Witten model---while
the requirement of even $x^4$-parity does not rule out the exotic mode $\G$, some further projection
may be appropriate for the pure-glue case.}
This would also be in line with the fact that the gluon condensate of the Witten model, Eq.~(\ref{gluoncondensate}),
is small, close to its standard SVZ value \cite{Shifman:1978bx},
while models with only one scalar glueball field
\cite{Ellis:1984jv,Janowski:2014ppa} cannot reconcile a small gluon
condensate with narrow glueball states.

In the range 1.5-1.8 GeV, where lattice gauge theory locates the lowest scalar glueball,
there are, experimentally, two isoscalar mesons $f_0(1500)$ and $f_0(1710)$ which are
frequently and alternatingly considered as predominantly glue.
The experimental results for the decay width into two pions are
\bea
&&\Gamma^{\rm (ex)}(f_0(1500)\to\pi\pi)/(1505{\rm MeV})=0.025(3),\\
&&\Gamma^{\rm (ex)}(f_0(1710)\to\pi\pi)/(1722{\rm MeV})= \left\{0.017(4) \atop 0.009(2) \right.
\eea
where the first result is taken from Ref.~\PDG, 
the second from 
Ref.~\cite{1208.0204} using data from the BES collaboration
\cite{hep-ex/0603048} (upper entry) and the WA102 collaboration \cite{hep-ex/9907055} (lower entry), respectively.

The lowest (exotic) scalar glueball mode $\G$ appears to have a much too large decay width to
be consistent with a dominantly glueball interpretation of either $f_0(1500)$ or $f_0(1710)$. 
On the other hand, the dilatonic mode has a decay width
below but comparable to the data for the two glueball candidates; in the case of the WA102 data
for the $f_0(1710)$ there happens to be even complete agreement.
In order to get a more complete picture, we shall now consider also the other
couplings between glueballs and mesons as determined by the Witten-Sakai-Sugimoto model.

\subsection{Glueball decay to four and more pions}

\begin{figure}[t]
\includegraphics[width=0.7\textwidth]{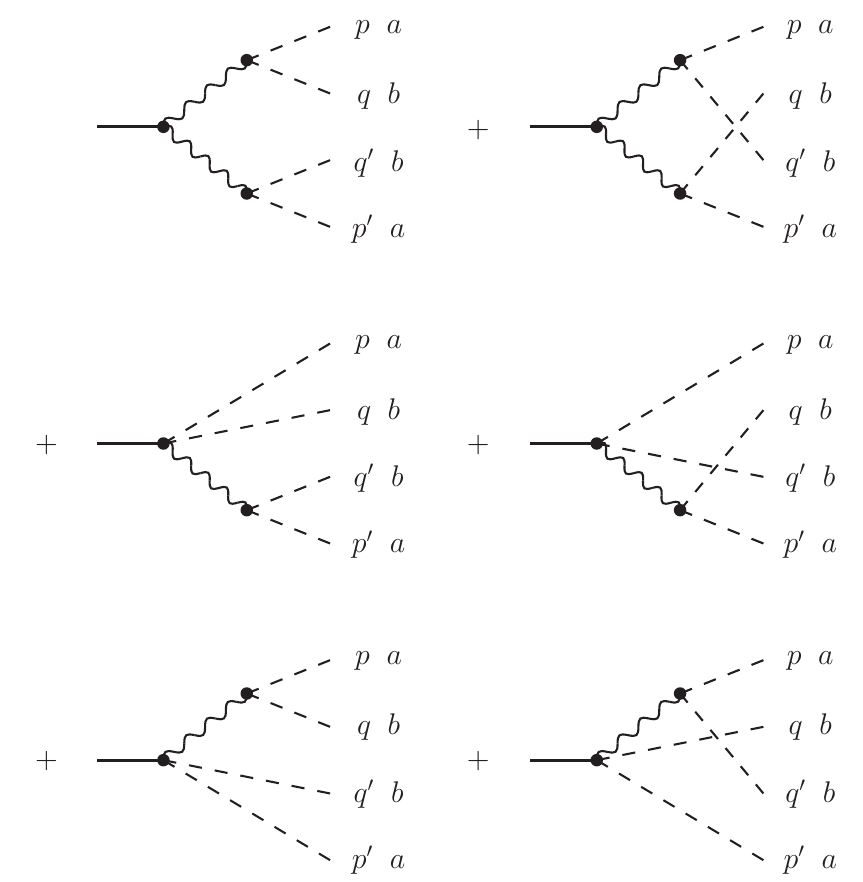}
\caption{Leading-order glueball decay into four pions, isospin indices $a\not=b$.}
\label{fig:g4pi}
\end{figure}

To leading order in $1/\alpha'$ or equivalently inverse 't Hooft coupling,
the D8 brane action (\ref{SD8F2}) does not give direct couplings of
glueballs to more than two pions. These appear only through
higher DBI corrections with terms quartic in the field strength $F_{\mu Z}$
as will be discussed further below.

Decays into more than four pions can however proceed through vertices involving
vector mesons.
The vertices coupling a single glueball to $\pi$ and/or $\rho$ mesons that are obtained
from the Yang-Mills part of the D8 brane action (\ref{SD8F2}) arise from terms of the form
(dropping derivatives and Lorentz indices)
\be\label{GverticesLO}
G\Tr(\pi\pi), \quad G\Tr(\rho\rho), \quad G\Tr(\rho[\pi,\pi]), \quad G\Tr([\pi,\rho]^2),
\quad G\Tr(\rho[\rho,\rho]),\quad G\Tr([\rho,\rho]^2).
\ee
Only the first three couplings are relevant for the decay of a glueball to $\le4$ pions.
The corresponding interaction Lagrangians for the exotic and the dilatonic scalar glueball
are given explicitly in Appendix \ref{sec:Lglueint} with the coupling constants for the lowest glueball states
listed in Table \ref{tabcG} and \ref{tabct}.

The relative width of the decay of a glueball to two pions was found above to be
$\Gamma_{G\to\pi\pi}/M \propto \lambda^{-1}N_c^{-2}$, parametrically suppressed
by a factor $1/N_c$ compared to the decay of the $\rho$ meson.
For glueballs with mass larger than $2m_\rho$, the decay into two $\rho$ mesons
is of the same parametric order. However, both the lowest exotic glueball and the lowest dilatonic
glueball have mass below the $2\rho$ threshold. In this case at least one $\rho$ meson
has to be off-shell, which leads to an additional suppression by a factor $\Gamma_\rho/m_\rho\propto \lambda^{-1}N_c^{-1}$.

Because the vertex coupling a single $\rho$ meson to two pions involves $\Tr(\rho[\pi,\pi])$,
the leading-order decay into four pions produces pairs of pions with different isospin index.
(The parametrically suppressed
decay $G\to2G\to4\pi^0$ and $G\to G+2\pi^0\to4\pi^0$ which only needs the leading Yang-Mills part
of the DBI action will be discussed together with the direct
decay $G\to4\pi^0$ from higher-order DBI corrections further below.)

\subsubsection{Leading-order decay rate of scalar glueballs to four pions involving $\pi^\pm$}

The Feynman diagrams for the amplitude of the decay $G_{E,D}\to 2\pi^a+2\pi^b$ with $a\not=b$
are shown in Fig.~\ref{fig:g4pi}. 
Some details of the rather lengthy calculation of the decay rate are given in Appendix \ref{sec:phint}.

Because one internal $\rho$ meson can reach its mass shell,
while it has nonnegligible width, we include (following Ref.~\cite{Hashimoto:2007ze})
$\Gamma_\rho$ in the $\rho$ meson propagator according to
$\Delta_\rho(r)=1/(r_0^2-\mathbf r^2-m_\rho^2+im_\rho \Gamma_\rho)$ with $\Gamma_\rho$ given by
(\ref{Gammarho}). This corresponds to a partial summation of higher-order terms in
inverse powers of $\lambda N_c$. As a crosscheck of our calculations, 
we have verified that in the limit $\lambda N_c\to\infty$
the resulting decay rate agrees with the rate for $G_{E,D}\to \rho\pi\pi$,
and in the case of glueballs above the $2\rho$ threshold, with $G\to2\rho$ (Appendix \ref{sec:Grhopipi}).

Because $m_\rho<m_{E,D}<2 m_\rho$, the leading parametric order of the decay width of $\G$ and $\D$ into four pions
is given by the process $G\to\rho\pi\pi$ and reads $\lambda^{-2}N_c^{-3}$. Decays through off-shell
$\rho$ mesons contribute terms of order $\lambda^{-3}N_c^{-4}$. 

For $\G$, which is only 10\% heavier
than a $\rho$ meson, the contribution from one on-shell $\rho$ meson is strongly suppressed by phase space, but the finite
width of the $\rho$ meson helps to increase the rate. For $\lambda\approx 16.63$ we
find\footnote{
Omitting the contributions from the interaction
terms involving the $\breve c$ coefficients as in \cite{Hashimoto:2007ze}
would give the even lower value $5.1\times 10^{-5}$.
In contrast to the decay into 2 pions, in the 4-pion decay rates the factors $\sqrt2$
in the coupling constant $g_{\rm YM}$ and in the normalization of the lowest scalar glueball
by which we differ from Ref.\ \cite{Hashimoto:2007ze}
no longer cancel. However, even when using exactly the couplings of Ref.~\cite{Hashimoto:2007ze}
we have not been able to reproduce the numerical
result $2.2\times 10^{-5}$
given in Eq.~(3.26) of \cite{Hashimoto:2007ze}.
}
\be
\Gamma_{\G\to4\pi}/M_E \approx 1.33\times 10^{-4} \quad (\lambda\approx 16.63). 
\ee

For the heavier dilatonic glueball $\D$, the process $G\to\rho\pi\pi$ is more dominant, leading to a significantly
larger relative width
\be
\Gamma_{\D\to4\pi}/M_D \approx 2.44\times 10^{-3} \quad (\lambda\approx 16.63). 
\ee
Evidently, the $4\pi$ decay of the lowest holographic glueball state, be it $\G$ or $\D$, is
strongly suppressed. Table \ref{tab4pi} summarizes these results and also shows
them for smaller $\lambda=12.55$.
(In Section \ref{sec:extrapol} we shall consider the extrapolation of
these lowest states to the higher masses of experimental glueball candidates in the range
predicted by lattice gauge theory.)

\subsubsection{Decay of excited scalar glueballs to two vector mesons}

For the excited dilatonic glueball with mass $M_{D^*}\approx 2358.4$~MeV, which
is above the $2\rho$ threshold,
a similar calculation, but with coefficients $d^*_i$ in place of $d_i$ (see Table \ref{tabct}),
gives 
\be\label{Ds4pi}
\Gamma_{\Ds\to4\pi}/M_{D^*} \approx 0.104 \quad (\lambda\approx 16.63).
\ee
This result, which involves resummed $\rho$ propagators, is in fact well approximated by
the decay rate to two on-shell $\rho$ mesons:
\be\label{Ds2rho}
\Gamma_{\Ds\to2\rho}/M_{D^*} \approx \frac{14.330}{\lambda N_c^2} \approx 0.096 \quad (\lambda\approx 16.63),
\ee
which corresponds to the strictly leading-order part of (\ref{Ds4pi}) as explained in Appendix \ref{sec:Grhopipi}.

The result (\ref{Ds2rho}), divided by its isospin factor of 3, also gives the decay into two isosinglet
vector mesons $\omega$, whose mass is only 1\% higher than that of the $\rho$ meson.

Since the decay width into two pions given in (\ref{Ds2piwidth}) is much smaller than the width into two vector mesons,
the excited dilatonic glueball turns out to decay predominantly into four pions and six pions.

The excited exotic scalar (if we do not discard this mode altogether) is instead dominated by the decay into two pions, which makes this state extremely broad. Calculating also the decay into two vector mesons, we find that the decay into two $\rho$ mesons
accounts for only about a third of the total decay into four pions, 
\be\label{Es2rho}
\Gamma_{\Gs\to2\rho}/M_{E^*} \approx \frac{2.078}{\lambda N_c^2} \approx 0.014 \quad (\lambda\approx 16.63).
\ee
This means that the decay into four pions is coming largely from the
$\Gs\rho\pi\pi$ vertex.

In Table \ref{tabexcsc} the results for the decay widths of the excited exotic and dilatonic scalar glueballs
in the Witten-Sakai-Sugimoto model are summarized. 
While the excited dilatonic scalar glueball has a 
more moderate decay width compared to the very broad excited exotic scalar, it turns out to be
still quite large, around 500 MeV.

\begin{table}
\begin{tabular}{l|r|c}
\toprule
& $M$ & $\Gamma/M$ \\ 
\colrule
$\G\to4\pi$ & 855 & $1.3\times10^{-4}$ \ldots $3.0\times10^{-4}$ \\ 
\colrule
$\D\to4\pi$ & 1487 & $2.4\times10^{-3}$ \ldots $3.9\times10^{-3}$ \\ 
$\D\to4\pi^0$ (NLO-DBI) & 1487 & $4.0\times10^{-6}$ \ldots $2.9\times10^{-5}$\\ 
$\D\to \G+2\pi^0\to4\pi^0$ & 1487 & $2.6\times 10^{-6}$ \ldots $4.5\times 10^{-6}$\\
$\D\to \D+2\pi^0\to4\pi^0$ & 1487 & $1.9\times 10^{-9}$ \ldots $4.5\times 10^{-9}$\\
\botrule
\end{tabular}
\caption{Decay widths of lowest exotic and lowest dilatonic scalar glueballs into four (massless) pions divided by glueball mass for $\lambda=16.63\ldots12.55$.}
\label{tab4pi}
\end{table}

\begin{table}
\begin{tabular}{l|r|c}
\toprule
& $M$ & $\Gamma/M$ \\ 
\colrule
$\Gs\to\{2\pi,2K,2\eta\}$ & 2168 & 0.397\ldots0.526\\
$\Gs\to4\pi$ & 2168 & 0.037\ldots0.061 \\ 
$\Gs\to2\omega\to6\pi$ & 2168 & 0.005\ldots0.006 \\
$\Gs\to2\phi$ & 2168 & 0.005\ldots0.006 \\ 
$\Gs$ (total) & 2168 & 0.443\ldots0.599\\
\colrule
$\Ds\to4\pi$ & 2358 & 0.104\ldots0.142 \\ 
$\Ds\to2\omega\to6\pi$ & 2358 & 0.032\ldots0.043 \\
$\Ds\to2\phi$ & 2358 & 0.032\ldots0.043 \\
$\Ds\to\{2\pi,2K,2\eta\}$ & 2358 & 0.029\ldots0.039\\
$\Ds$ (total) & 2358 & 0.197\ldots0.267\\
\botrule
\end{tabular}
\caption{Decay widths of excited scalar glueballs divided by glueball mass for $\lambda=16.63\ldots12.55$ (chiral limit, with a ratio 3:4:1 for the combined decay into $2\pi,2K,2\eta$).}
\label{tabexcsc}
\end{table}

\subsubsection{Scalar glueball decay to four $\pi^0$}
\begin{figure}[t]
\centerline{\includegraphics[width=0.3\textwidth]{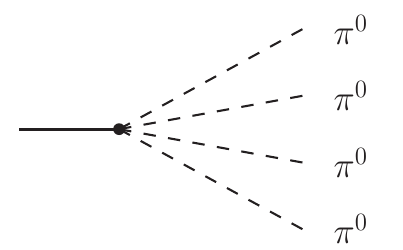}}
\caption{Glueball decay into four $\pi^0$ through a vertex from the next-to-leading
order terms of the DBI action;}
\label{fig:g4pi0}
\end{figure}

The glueball decays into four pions
that we have considered above involve pairs of pions with different isospin index.
A decay to four $\pi^0$ is suppressed by powers of inverse 't Hooft coupling, because it
either has to come from higher-order contributions in the DBI action of the D8 branes (Fig.~\ref{fig:g4pi0})
or has to involve glueball self-interactions and virtual glueballs (Fig.~\ref{fig:gg4pi0}).

As shown in Appendix \ref{sec:Lglueint}, the parametric order of the vertex
formed by a single glueball and four $\pi^0$ turns out to be $\lambda^{-7/2}N_c^{-2}$,
whereas the amplitude for $G\to2G\to4\pi^0$ and $G\to G+2\pi^0\to4\pi^0$
is proportional to $\lambda^{-3/2}N_c^{-3}$.
The former thus has stronger suppression in inverse powers of $\lambda$, while
the latter is more strongly suppressed with respect to inverse powers of $N_c$.

For simplicity, we only consider the
dilatonic glueball, since the exotic glueball has a much more complicated interaction
Lagrangian. In Appendix \ref{sec:LD4pi0} the interaction Lagrangian for a dilatonic glueball
with four $\pi^0$ resulting from the next-to-leading terms of the DBI action
has been obtained, and in Appendix \ref{sec:LDD2pi0} the vertex for $\D\to G_{D,E}+2\pi^0$.
Numerically evaluating the respective decay rates of the dilatonic glueball
shows that at finite 't Hooft coupling and $N_c=3$ the dominant decay process
comes from the direct coupling of $\D$ to four $\pi^0$. For $\lambda\approx 16.63$
we find (see Appendix \ref{sec:GDto4pi0} for details)
\be\label{GammaDto4pi0}
\Gamma^{\rm (NLO-DBI)}_{G_D\to4\pi^0}/M_D\approx 4.02\times 10^{-6} \quad (\lambda\approx 16.63).
\ee

The decay through virtual glueballs, while not as strongly suppressed by inverse powers
of $\lambda$, is subleading at large $N_c$ and is disfavored by phase space.
To check whether it might nevertheless be important at $N_c=3$ and our range
of 't Hooft coupling, we have evaluated the first diagram in Fig.~\ref{fig:gg4pi0}
involving one virtual glueball and found that its contribution is smaller than (\ref{GammaDto4pi0})
by
several orders of magnitude (see Table \ref{tab4pi}), 
\be
\Gamma^{\rm (LO-DBI)}_{\D\to\D+2\pi^0\to4\pi^0}/M_D\approx 1.94\times 10^{-9} \quad (\lambda\approx 16.63).
\ee
If we do not discard the exotic glueball as a physical state
(for instance if we were to interpret the latter as holographic dual of
a glueball component of the $\sigma$-meson, as speculated at the end of Section~\ref{sec:G2pi}),
we should also 
consider the process $\D\to\G+2\pi^0$ which is less suppressed kinematically
(but still by $N_c^{-1}$). This would be of similar magnitude as
the result (\ref{GammaDto4pi0}):\footnote{By contrast, in the scenario of Ref.~\cite{Narison:1996fm}, where the $\sigma$-meson has
a large glue contribution,
the heavier glueball is claimed to have important $4\pi^0$ decays.}
\be\label{GDGE00}
\Gamma^{\rm (LO-DBI)}_{\D\to\G+2\pi^0\to4\pi^0}/M_D\approx 2.56\times 10^{-6} 
\quad (\lambda\approx 16.63).
\ee
(As shown in Table \ref{tab4pi}, at smaller $\lambda$ 
this contribution is less important compared to
the next-to-leading DBI contribution (\ref{GammaDto4pi0}).)

\begin{figure}[t]
\centerline{\includegraphics[height=0.23\textwidth]{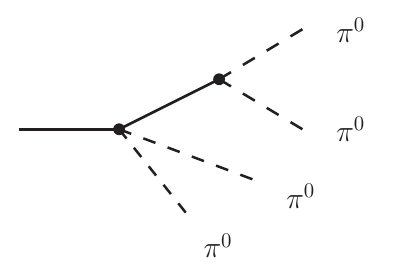}\includegraphics[height=0.23\textwidth]{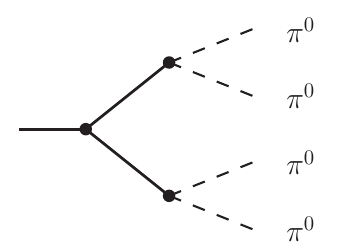}}
\centerline{\hfil (a) \hfil\hfil (b) \hfil}
\caption{Glueball decay into four $\pi^0$ 
(a) through terms in the Yang-Mills part of the DBI action
that are quadratic in the glueball mode;
(b) through a pair of virtual glueballs.}
\label{fig:gg4pi0}
\end{figure}

Decays into four $\pi^0$ have been seen for the glueball candidate $f_0(1500)$
at a level of about an order of magnitude below the general $4\pi$ decay \PDG,
whereas no such data seem to be available for $f_0(1710)$.
The smallness of the holographic result (\ref{GammaDto4pi0}) however would correspond to a much stronger
suppression than the one observed experimentally for 
$f_0(1500)$.

\subsubsection{Tensor glueball decay to two vector mesons}\label{sec:exctensordecay}

Unless the mass of the lowest tensor glueball is manually adjusted (as we shall
consider to do in Section \ref{sec:extrapol}), only the excited tensor glueball
of the Witten model with mass $M_{T^*}=M_{D^*}\approx 2358.4$~MeV can decay into two $\rho$ or two $\omega$ mesons.

The decay rate involves two sums over the polarizations of the two vector mesons.
The average over the polarization of the tensor can again be performed by choosing
the particular polarization $\epsilon^{11}=-\epsilon^{22}=1$ and averaging over spatial directions. The rate for two vector mesons with fixed isospin quantum number reads
\be\label{Tstorhorho}
\Gamma=\frac{1}{16\pi M_{T^*}} \sqrt{(M_{T^*}/2)^2-m_\rho^2}
\int\frac{d\Omega}{4\pi}\sum_{\lambda_1,\lambda_2=1}^3 |\mathcal M_{\epsilon_T}(\lambda_1,\lambda_2)|^2,
\ee
where $\lambda_{1,2}$ are labels for the polarizations of the two vector mesons and
$\epsilon_T$ refers to the specific tensor polarization.
The amplitude $\mathcal M_{\epsilon_T}(\lambda_1,\lambda_2)$ and
the final result of the summations and the integration are given in Appendix \ref{sec:MT2rho}.
With coupling constants $t_i^*\equiv\sqrt6 d_i^*$ and $d_i^*$ from Table
\ref{tabct}, the result for the decay into two $\rho$ mesons is
\be\label{Tsrhorho}
\frac{\Gamma_{T^*\to2\rho\to4\pi}}{M_{T^*}}\approx\frac{21.236}{\lambda N_c^2}\approx 0.142\quad (\lambda\approx 16.63),
\ee
and $1/3$ of that result for the decay $T^*\to2\omega\to6\pi$.
This should be compared to the decay rate into two pions, Eq.~(\ref{Tstopipi}), which
is less than $1/8$ of 
(\ref{Tsrhorho}).

As we shall discuss below, a similar pattern arises when the lowest tensor
glueball is extrapolated in mass such that it is above the $2\rho$ threshold.

\section{Extrapolations and comparison with experimental data}\label{sec:extrapol}

When comparing our results for decay rates with experiment, it seems reasonable to
do so with the dimensionless ratio $\Gamma/M$ when extrapolating the mass $M$
of the holographic glueball to the mass of the experimental glueball candidates $f_0(1500)$
or $f_0(1710)$. In the case of decay into two massless pions, Eqs.~(\ref{GEtopipi})--(\ref{Tstopipi}),
this ratio involves two explicit powers of the glueball mass $M$ that cancel the inverse mass scale squared
coming from the normalization of the glueball field, Eq.~(\ref{Hnorm}) or (\ref{HDTnorm}).
When extrapolating to higher glueball masses, we thus assume that the normalization of the glueball field
scales according to the glueball mass. While this keeps $\Gamma/M$ for two-pion decays unchanged,
the decay rates into two vector mesons or four pions are modified and
depend in fact strongly on whether the glueball mass is above or below the $2\rho$ threshold.

\subsection{Extrapolations for the scalar glueball candidates $f_0(1500)$
and $f_0(1710)$}

The results of such an extrapolation to the experimental masses of the isoscalar
mesons $f_0(1500)$
or $f_0(1710)$ is given
in Table \ref{tab4piexp}, where the holographic results of the
(chiral) Witten-Sakai-Sugimoto model for the lowest (``exotic'') and the dilatonic $0^{++}$
glueball are
compared to the experimental results for the total and the partial decay widths.
Here we have generalized our results to $N_f=3$ and assumed that pions, kaons and $\eta$ mesons
appear in ratios $3:4:1$, respecting SU(3) flavor symmetry. 

Explicit masses for quarks would require a modification of the Sakai-Sugimoto model, for example
along the lines of Ref.~\cite{Aharony:2008an,Hashimoto:2008sr,McNees:2008km}, which we intend to study in future work.
This will necessarily modify the coupling of scalar glueballs through contributions that depend on
the mass of the pseudo-Goldstone bosons, and this may either increase or decrease
the decay amplitudes into the heavier pseudo-Goldstone bosons.
A significant enhancement would be in line with the so-called chiral suppression of scalar glueball decays
that is suggested by the lattice results of
Ref.~\cite{Sexton:1995kd} and the analysis of Ref.~\cite{Chanowitz:2005du}.
(In the dilaton
effective theory of Ref.~\cite{Ellis:1984jv} also an increase of
the amplitude for the decay into a pair of heavier pseudo-Goldstone bosons was found, however
such that it is approximately canceled
by the kinematical suppression from the phase space integral.)

When comparing the extrapolated decay rates of the holographic glueballs with those
of the isoscalar mesons $f_0(1500)$ or $f_0(1710)$ we find that the lowest (exotic) glueball
is much too broad to be identified as their dominant glueball component.
The dilatonic glueball, however, is sufficiently
narrow for this purpose. It leads to a total decay width that is quite close to the experimental
width of $f_0(1710)$, while being somewhat more strongly below that of $f_0(1500)$.
With mass equal to that of $f_0(1500)$, the dilatonic glueball has significantly smaller width in
$2\pi$ decays, and still smaller for decays into four pions, which is the dominant decay mode
of the $f_0(1500)$. 

Regarding the $f_0(1710)$, the decay into $2\pi$ is found to
be nicely comparable to the experimental value, 
while the stronger rate into pairs of heavier pseudo-Goldstone bosons 
remains unaccounted for with our assumption of SU(3) invariance.
A significant enhancement of decays into kaons and $\eta$ mesons may however be
brought about by mass terms for the latter which inevitably will give additional contributions
to the coupling with scalar glueballs.

If our extrapolation of the decay width into $4\pi$ can be trusted, this appears now
uncomfortably large considering that the decay of $f_0(1710)$ into $4\pi$ has not been observed. It should be
noted, however, that the experimental data for the branching ratios of the $f_0(1710)$ still have
large uncertainties and are not covered by the Particle Data Group \PDG. The quoted results are
from Refs.~\cite{1208.0204,Janowski:2014ppa}, which assume that decays into $\pi\pi$, $\eta\eta$, and $K \bar K$
add up to the total width with negligible contribution from $4\pi$ decays.

Our extrapolations also predict decays into two $\omega$ mesons at a nonnegligible level.
According to \PDG, decays of $f_0(1710)$ to two $\omega$ mesons have at least been seen.
The Witten-Sakai-Sugimoto model for a (pure) glueball candidate suggests that the rate into four pions
should be about twice as large.


\begin{table}
\begin{tabular}{l|r|r||r|r}
\toprule
decay & $M^{\rm exp}$ & $\Gamma/M$ (exp.) & $\Gamma/M[\G({M^{\rm exp}})]$ & $\Gamma/M[\D({M^{\rm exp}})]$ \\
\colrule
$f_0(1500)$ (total) & 1505 & 0.072(5) & 0.249\ldots0.332 & 0.027\ldots0.037 \\
$f_0(1500)\to4\pi$ & 1505 & 0.036(3) & \it 0.003\ldots0.006 &  0.003\ldots 0.005 \\
$f_0(1500)\to2\pi$ & 1505 & 0.025(2) & 0.092\ldots0.122 & 0.009\ldots0.012\\
$f_0(1500)\to 2K$ & 1505& 0.006(1) & 0.123\ldots0.163 & 0.012\ldots0.016 \\
$f_0(1500)\to 2\eta$ & 1505& 0.004(1) & 0.031\ldots0.041 & 0.003\ldots0.004 \\
\colrule
$f_0(1710)$ (total) & 1722 & 0.078(4) & 0.252\ldots0.336 & \it 0.059\ldots0.076 \\[4pt]
$f_0(1710)\to 2K$ & 1722& * $\left\{ 0.041(20) \atop 0.047(17) \right.$& 0.123\ldots0.163 & 0.012\ldots0.016 \\[12pt]
$f_0(1710)\to 2\eta$ & 1722& * $\left\{0.020(10) \atop 0.022(11) \right.$ & 0.031\ldots0.041 & 0.003\ldots0.004 \\[12pt]
$f_0(1710)\to2\pi$ & 1722 & * $\left\{0.017(4) \atop 0.009(2) \right.$ & 0.092\ldots0.122 & 0.009\ldots0.012 \\[8pt]
$f_0(1710)\to4\pi$ & 1722 & ? & \it 0.006\ldots 0.010 & \it 0.024\ldots 0.030 \\ 
$f_0(1710)\to2\omega\to6\pi$ & 1722 & seen & \it 0.00016\ldots0.00021 & \it 0.011\ldots 0.014 \\ 
\botrule
\end{tabular}
\caption{Experimental data for the decay rates of the isoscalar mesons $f_0(1500)$ and $f_0(1710)$
juxtaposed to 
the holographic results for the various decay channels
of the lowest (exotic) glueball ($G_E$) and predominantly dilatonic glueball ($G_D$) 
with mass $m_{E,D}$  artificially raised to the respective experimental values 
(still in the chiral limit, i.e.\ with massless pions, kaons, and $\eta$) and 't~Hooft coupling
varied from 16.63 to 12.55. Experimental data are from Ref.~\PDG\ except for those marked by a star, which
are from Ref.~\cite{1208.0204} where the total width of $f_0(1710)$ was split under the assumption of
a negligible branching ratio to four pions,
using data from BES \cite{hep-ex/0603048} (upper entry) and WA102 \cite{hep-ex/9907055} (lower entry),
respectively.
(Holographic predictions that are substantially increased due to the manually adjusted
glueball mass are rendered in italics.)
}
\label{tab4piexp}
\end{table}

In this context it is worth mentioning that there are still many open questions
surrounding the nature of $f_0(1710)$ \cite{Klempt:2007cp}. For example, some authors have argued
that the nearby resonance \cite{Ablikim:2004wn} $f_0(1790)$, which is not yet covered by the Particle
Data Group, should be combined with $f_0(1710)$ into one object $f_0(1760)$, for which
Ref.~\cite{Anisovich:2002ij} was able to fit disparate decay patterns, with and without
significant decay into four pions.

\subsection{Extrapolations for the tensor glueball}

In the Witten model, with $\MKK=949$~MeV, the mass of the tensor glueball equals the mass of
the dilatonic scalar glueball, and the tensor glueball has roughly similar decay rates into two and four pions.
The rate into two pions practically exhausts the decays into pions, and
has been calculated above in Eq.~(\ref{Ttopipi}). The lowest tensor glueball
thus turns out to be a rather narrow state, however this is due to the fact that
it stays below the $2\rho$ threshold.

Indeed, the situation is markedly different for the excited tensor glueball $T^*$.
Its mass equals that of the excited dilatonic glueball, and
because this is above the threshold for two $\rho$ mesons, there is a significant contribution to
four-pion decays, and also from other vector meson decays, as we have seen in Section \ref{sec:exctensordecay}.
Extrapolating the couplings of the lowest tensor glueball to a similarly high mass,
2 or 2.4~GeV (where the latter is roughly the prediction of lattice gauge theory for the lowest tensor mode), equally
gives large contributions from decay into two vector mesons, as listed
in Table \ref{tabtensorextrapolmpi}. Reassuringly, these results
are quite close to those for the unmodified results for $T^*$, cp.~Eq.~(\ref{Tsrhorho}), so that we consider them
as plausible extrapolations to the likely situation of a tensor glueball with mass above 2 GeV.

\begin{table}
\begin{tabular}{l|r|r}
\toprule
decay & M & $\Gamma/M[T(M)]$ \\
\colrule
$T\to 2\pi$ & 1487 & 0.013\ldots0.018 \\
$T\to 2K$ & 1487 & 0.004\ldots0.006 \\
$T\to 2\eta$ & 1487 & 0.0005\ldots0.0007 \\
$T$ (total) & 1487 & $\approx 0.02\ldots0.03$\\
\colrule
$T\to 2\rho\to 4\pi$ & 2000 & 0.135\ldots 0.178 \\ %
$T\to 2K^*\to 2(K\pi)$ & 2000 & 0.119\ldots 0.177 \\
$T\to 2\omega\to 6\pi$ & 2000 & 0.045\ldots 0.059 \\ 
$T\to 2\pi$ & 2000 & 0.014\ldots0.018 \\
$T\to 2K$ & 2000 & 0.010\ldots0.013 \\
$T\to 2\eta$ & 2000 & 0.0018\ldots0.0024 \\
$T$ (total) & 2000 & $\approx 0.32\ldots0.45$\\
\colrule
$T\to 2K^*\to 2(K\pi)$ & 2400 & 0.173\ldots 0.250 \\
$T\to 2\rho\to 4\pi$ & 2400 & 0.159\ldots 0.211 \\ %
$T\to 2\omega\to 6\pi$ & 2400 & 0.053\ldots 0.070 \\ 
$T\to 2\phi$ & 2400 & 0.032\ldots 0.051 \\ 
$T\to 2\pi$ & 2400 & 0.014\ldots0.019 \\
$T\to 2K$ & 2400 & 0.012\ldots0.016 \\
$T\to 2\eta$ & 2400 & 0.0025\ldots0.0034 \\
$T$ (total) & 2400 & $\approx 0.45\ldots0.62$\\
\botrule
\end{tabular}
\caption{Extrapolation of tensor glueball decay for the case of massive pseudo-Goldstone bosons with glueball mass
$M=M_T=M_D$ and when the latter is raised to 2 GeV or
the lattice prediction $\sim 2.4 \,{\rm GeV}$.
The 't~Hooft coupling is again
varied from 16.63 to 12.55.
}
\label{tabtensorextrapolmpi}
\end{table}

In Table \ref{tabtensorextrapolmpi} we have also extrapolated to decays into kaons and $\eta$ mesons.
In the holographic setup,
a tensor glueball presumably does not couple to an explicit mass term of the pseudoscalar mesons,
so the effect of the latter should be purely kinematic. The results (\ref{MTpipi}) and
(\ref{GTpipi}) imply a pseudoscalar mass dependence of the form $(1-4m_\pi^2/M_T^2)^{5/2}$.
This suppression is such that it overcompensates the ratio 4/3 that favors kaons over pions.

For the decays into vector mesons $K^*$ and $\phi$ we have taken into account that their masses
are larger than $m_\rho$ in the phase space factor, but we have left open the possibility that
this also increases the coupling $t_2\equiv \sqrt6 d_2$ and merged the two alternatives
in the range of results for the corresponding decay rates for tensor glueballs with
increased mass.

Adding up the individual contributions, we find a very broad width for a 2.4 GeV tensor glueball, 1.1 to 1.5 GeV, which is much broader than all the $f_2$ mesons listed in \PDG. With a mass around 2 GeV, the width (600 to 900 MeV)
turns out to be larger but perhaps marginally comparable with that of the tensor meson $f_2(1950)$, which has $\Gamma= 472(18)$ MeV.
The latter is indeed occasionally discussed as a candidate for a tensor glueball as it
appears to have largely flavor-blind decay modes.

\section{Conclusion}

Using the Witten-Sakai-Sugimoto model for holographic QCD, which
only has one free dimensionless parameter, 
we have repeated and extended
the calculation of glueball decay rates of Ref.~\cite{Hashimoto:2007ze},
where only the lowest scalar mode was studied.

This lowest mode is associated with an exotic polarization of
the gravitational field, involving components in the direction of
compactification from a 5-dimensional super-Yang-Mills theory down to
nonsupersymmetric Yang-Mills theory. 
The mass of this lowest (``exotic'') $0^{++}$ mode turns out to be only
slightly above the mass of the $\rho$ meson and is therefore
much smaller than the mass scale of glueballs found in lattice gauge theory.
The background of the Witten model also contains another
tower of scalar glueball modes which are predominantly dilatonic
and whose lowest mass is about 1.5 GeV, not far from the predictions
of lattice simulations.

Besides its very low mass,
the lowest (exotic) scalar glueball turns out to have a decay rate that is significantly higher
than that of the heavier dilatonic mode, which seems counterintuitive
if the latter were to represent an excitation of the former.
We are therefore led to the conjecture that the exotic scalar mode should be
discarded so that the glueball spectrum begins with the
(predominantly) dilatonic mode as lowest glueball.
Another, more speculative possibility that we have mentioned in
Section \ref{sec:G2pi} is that the exotic scalar mode
represents a broad glueball component of the $\sigma$-meson in line
with the scenario of Ref.~\cite{Narison:1996fm,Narison:2005wc,Kaminski:2009qg},
which features a broad glueball around 1 GeV and a narrower one around 1.5 GeV.

The decay widths of glueballs obtained in the Witten-Sakai-Sugimoto model
are parametrically suppressed by a factor of $\lambda^{-1}N_c^{-2}$, but
the numerical results vary substantially for the different modes and decay channels,
and thus do not give a picture of ``universal narrowness'' despite
the large-$N_c$ nature of the Witten-Sakai-Sugimoto model.

A very strong parametric suppression is obtained for the decay into $4\pi^0$,
as already pointed out in Ref.~\cite{Hashimoto:2007ze}. We have confirmed
that also the final numerical value turns out to be very small.

A noteworthy feature of the Witten-Sakai-Sugimoto model is that
the value of the gluon condensate is small, close to its standard SVZ value \cite{Shifman:1978bx},
whereas phenomenological models which incorporate a scalar glueball through
a QCD dilaton field \cite{Ellis:1984jv,Janowski:2014ppa} would require
very large gluon condensates to admit only narrow glueball states.

We have also extrapolated our results so that they can be
compared with experimental data for the scalar glueball candidates
$f_0(1500)$ or $f_0(1710)$. In the case of $f_0(1500)$,
our results for the decay widths of the dilatonic glueball are significantly
below the observed rates for decay into two pions and even more so for the
experimentally dominant decay into four pions.
In the case of the $f_0(1710)$ meson, the decay rate into two pions
comes out in nice agreement with available experimental data. The much stronger rate into kaons
is not accounted for, but this may be due to the fact that the Witten-Sakai-Sugimoto model 
is strictly chiral and the mechanism of chiral suppression \cite{Sexton:1995kd,Chanowitz:2005du}.
However our (crude) extrapolation to the mass of $f_0(1710)$
predicts also a significant branching ratio into four pions that
has not been seen experimentally. [Although in this context it should be noted that
the identification of
$f_0(1710)$ and its separation from the nearby $f_0(1790)$ \cite{Ablikim:2004wn} has been a matter of debate \cite{Anisovich:2002ij,Klempt:2007cp}].

Furthermore, we have studied the decay of tensor glueballs, which
in the Witten-Sakai-Sugimoto model have a narrow width into two pions
and (when the mass is above the $2\rho$ threshold) a large width into
four pseudoscalars, such that at best the isoscalar tensor meson $f_2(1950)$
appears to be (marginally) compatible with our the holographic result, while heavier
tensor glueballs would have to be broader than the tensor mesons
so far discussed in the literature.

In the case of the tensor glueball we can already plausibly anticipate
the effects of nonzero pseudo-Goldstone masses. In the case of scalar glueballs
the situation is less clear and we intend to study this issue in extensions
of the Witten-Sakai-Sugimoto model in
a future work. This would be particularly interesting in view of the glueball
candidate $f_0(1710)$ which according to Ref.~\cite{Janowski:2014ppa}
could be a nearly unmixed glueball and which has a ratio \PDG\ 
$\Gamma(\pi\pi)/\Gamma(KK)$ 
that is significantly below the flavor-symmetric value $3/4$. 

Since the holographic results pertain only to pure glueballs,
it would clearly be most interesting to study mixing of glueballs with
$q\bar q$ states as this can strongly obscure signatures of glueball
content.
In the holographic setup, mixing is suppressed by $1/N_c$  \cite{Hashimoto:2007ze}
and would presumably require more difficult stringy corrections
that are not captured by the effective Lagrangian following from
the Witten-Sakai-Sugimoto model.
Absent those, it might be interesting to consider a more phenomenological approach such as extended
linear sigma models \cite{Janowski:2014ppa}, where holographic results
for the glueball-meson interactions could be used as input instead of fitting to experimental data.

\begin{acknowledgments}
We would like to thank 
Koji Hashimoto, Chung-I Tan, and Seiji Terashima for correspondence and
David Bugg, Francesco Giacosa, Stanislaus Janowski, and Dirk Rischke for useful discussions. 
This work was supported by the Austrian Science
Fund FWF, project no. P26366, and the FWF doctoral program
Particles \& Interactions, project no. W1252.
\end{acknowledgments}

\appendix

\section{Ten-dimensional field equations}\label{sec:tendimfieldeq}

The Kaluza-Klein reduction of the eleven-dimensional graviton modes yields metric fluctuations pertaining to the four-sphere, i.e.\ in the components denoted by $g_{\Omega\Omega}$. Omitting these fluctuations as done in Ref.~\cite{Hashimoto:2007ze}
corresponds to dropping all vertices proportional to $\breve{c}_i$ in the interaction Lagrangian (\ref{LintGexotic}) of the exotic scalar glueball. To see whether this reduction could be justified, we check if these truncated modes solve the ten-dimensional field equations of type IIA supergravity. 

For the Witten-Sakai-Sugimoto model, the relevant terms of the supergravity action are given by \cite{Polchinski:1998rr}
\begin{equation}
S_{\mathrm{IIA}}=\frac{1}{2\kappa_{10}^2}\int d^{10}x\sqrt{-g}\mathrm{e}^{-2\Phi}\left(R+4\nabla_M\Phi\nabla^M\Phi-\frac{1}{2}\mathrm{e}^{2\Phi}|F_4|^2\right),
\end{equation}
where $2\kappa_{10}^2=(2\pi)^7l_s^8$ and $F_4=dC_3$ is the four-form from the R-R sector of the theory, with
\begin{equation}
|F_4|^2\equiv\frac{1}{4!}F_{ABCD}F^{ABCD}.
\end{equation}

Variation of this action with respect to the background metric $g_{MN}$ and the dilaton field $\Phi$ results in 
\begin{equation}
R_{MN}-\frac{1}{2}Rg_{MN}+2\nabla_M\Phi\nabla_N\Phi-\frac{\mathrm{e}^{2\Phi}}{4!}\left(2{F_M}^{ABC}F_{NABC}-3!|F_4|^2\right)=0
\end{equation}
and
\begin{equation}
R+4\nabla_M\nabla^M\Phi-4\nabla_M\Phi\nabla^M\Phi=0,
\end{equation}
respectively.

The solution to these equations that corresponds to the background of the Witten model is given by the metric (\ref{ds210}), the dilaton (\ref{Phibackground}) and a nonvanishing R-R four-form field. 
The latter is fixed by the requirement that the flux through a unit four-sphere is quantized, i.e.
\begin{equation}
\int_{S^4}F_4=N_c\frac{\kappa_{10}}{\sqrt{\alpha'\pi}},
\end{equation} 
where the factor of $N_c$ arises from the fact that we are considering a stack of $N_c$ D4-branes. The field strength that satisfies this condition is given by $F_4=3R_{\mathrm{D4}}g_s^{-1}\omega_4$, with $\omega_4$ denoting the volume form of the unit four-sphere.\footnote{Note that this result looks different in  some of the literature, e.g. \cite{Sakai:2004cn}. This is due to a different convention with rescaled three-form potential.}

With this information, one can linearize the field equations and plug in the solutions both with and without the spherical fluctuations, which is easily done with computer algebra tools. The result is that for both the dilatonic and exotic glueball modes, the field equations are not satisfied unless the 
fluctuations along the four-sphere are included. This means that in a rigorous top-down approach the vertices corresponding to the coefficients $\breve{c}_i$ have to be included in the calculation of decay rates.    
(For the dilatonic glueball mode, the need to include $g_{\Omega\Omega}$ fluctuations
can also be deduced from the explicit 10-dimensional calculations
of Ref.~\cite{Hashimoto:1998if}.)

\section{Glueball-meson interaction Lagrangians}\label{sec:Lglueint}

The effective interaction Lagrangian of glueballs and $q\bar q$ mesons is obtained
by inserting the 10-dimensional metric fluctuations (\ref{deltag10}) into the D8 brane
action and integrating over the bulk coordinates. In this section we give
the result for the lowest (exotic) scalar glueball, the dilatonic scalar glueball, and
the tensor glueball, expanded up to the order needed for the calculation of
decay rates of glueballs into two pions and four pions as discussed in the text.

As discussed above, we do so by keeping induced fluctuations in $g_{\Omega\Omega}$.
In the D8 brane action (\ref{SD8full})
the contribution from the dilaton fluctuation $\delta G_{11,11}$
appearing 
through the factor $\sqrt{g_{S^4}}=g_{\Omega\Omega}^2$ in $\sqrt{-\tilde g}$ 
is opposite in sign to that from $e^{-\Phi}$ and larger by a factor $4/3$.

Let us also recall that following Ref.~\cite{Hashimoto:2007ze} we use the convention
\be
\pi=\pi^a T^a,\quad
\rho_\mu=\rho^a_\mu T^a,\qquad
 \Tr T^a T^b=\delta^{ab},
\ee
so that for $N_f=2$ using Pauli matrices we have
$T^a=\sigma^a/\sqrt2$ and $\Tr [T^a,T^b]T^c=\sqrt2 i \epsilon^{abc}$,
while Ref.~\cite{Sakai:2004cn,Sakai:2005yt} have
$\Tr T^a T^b=\frac12\delta^{ab}$. The Minkowski metric used in the 3+1-dimensional 
Lagrangians is $\eta_{\mu\nu}={\rm diag}(-+++)$.

\subsection{Lowest scalar mode}

The glueball-meson interactions contributing at leading order to the decay of a glueball into two or four pions
are given by the terms linear in the glueball field and up to quadratic in $\rho$, maximally trilinear in $\pi,\rho$
in the Yang-Mills part of the DBI action of the D8 branes. In the case of the lowest (exotic) scalar mode,
they read
\bea\label{LintGexotic}
\mathcal L_{\G}&=-&\Tr\Biggl\{ 
c_1\left[\frac12\6_\mu\pi \6_\nu\pi
\frac{\6^\mu\6^\nu}{\MG^2}\G
+\frac14(\6_\mu\pi)^2\left(1-\frac{\Box}{\MG^2}\right)\G \right]\nonumber\\
&&+c_2\,\MKK^2
\left[\frac12 \rho_\mu\rho_\nu \frac{\6^\mu\6^\nu}{\MG^2}\G
+\frac14(\rho_\mu)^2\left(1-\frac{\Box}{\MG^2}\right)\G\right]\nonumber\\
&&+c_3
\left[\frac12 \bar F_{\mu\rho}\bar F_\nu{}^\rho \frac{\6^\mu\6^\nu}{\MG^2}\G
-\frac18 \bar F_{\mu\nu}\bar F^{\mu\nu}\left(1+\frac{\Box}{\MG^2}\right)\G \right]\nonumber\\
&&+c_4\frac3{2\MG^2} \rho_\mu \bar F^{\mu\nu}\6_\nu \G\nonumber\\
&&+ic_5 \left[\6_\mu\pi [\pi,\rho_\nu]\frac{\6^\mu\6^\nu}{\MG^2}\G
+\frac12 \6_\mu\pi [\pi,\rho^\mu]\left(1-\frac{\Box}{\MG^2}\right)\G
\right]\nonumber\\
&&+\frac12\breve c_1 \6_\mu\pi \6^\mu\pi \,\G
+\frac12\breve c_2 \,\MKK^2 \rho_\mu \rho^\mu \G\nonumber\\
&&+\frac14\breve c_3 \bar F_{\mu\nu}\bar F^{\mu\nu}\G 
+i\breve c_5\6_\mu\pi [\pi,\rho^\mu]\G
\Biggr\},
\eea
where $\bar F_{\mu\nu}\equiv \6_\mu\rho_\nu-\6_\nu\rho_\mu$ without
a commutator term $[\rho_\mu,\rho_\nu]$.
This agrees with Ref.~\cite{Hashimoto:2007ze}, whose notations we have adopted, in
the part involving $c_i$, but Ref.~\cite{Hashimoto:2007ze} effectively dropped all terms proportional to $\breve c_i$
due to the neglect of $\delta g_{\Omega\Omega}$.

The coefficients $c_i$ and $\breve c_i$ are obtained by integrals over the glueball
mode function $\HG(Z)$, the $\rho$ meson mode function $\psi_1(Z)$, and the pion
mode function $\phi_0(Z)\propto 1/K$, $K\equiv 1+Z^2$, according to
\bea\label{cbrevec}
c_1&=&\int dZ \frac{\bar \HG}{\pi K},
\qquad c_2=2\kappa\int dZ\,K(\psi_1')^2 \bar \HG,
\qquad c_3=2\kappa\int dZ\,K^{-1/3}(\psi_1)^2 \bar \HG,\nonumber\\
c_4&=&2\kappa\MKK^2\int dZ \frac{20 Z K}{(5K-2)^2}\psi_1\psi'_1 \HG,
\qquad c_5=\int dZ \frac{\psi_1\bar \HG}{\pi K},\nonumber\\
\breve c_1&=&\int dZ \frac{\HG}{4\pi K},
\qquad \breve c_2=\frac12\kappa\int dZ\,K(\psi_1')^2 \HG,\nonumber\\
\qquad \breve c_3&=&\frac12\kappa\int dZ\,K^{-1/3}(\psi_1)^2 \HG,
\qquad \breve c_5=\int dZ \frac{\psi_1 \HG}{4\pi K},
\eea
where the integral over $Z$ is from $-\infty$ to $+\infty$ and where 
following Ref.~\cite{Hashimoto:2007ze} we have introduced
\be
\bar \HG(Z)\equiv \left[\frac14+\frac3{5K-2}\right]\HG(Z).
\ee

The corresponding coefficients for the excited mode $\Gs$ are obtained by replacing
the lowest mode function $\HG(Z)$ by the next highest eigenfunction.

The numerical results for the coefficients $c_i$, $\breve c_i$ for the lowest mode
as well as for $c_1^*$ and $\breve c_1^*$ for $\Gs$ are given in Table \ref{tabcG}.

\subsection{Dilatonic and tensor mode}

\subsubsection{Glueball-meson interactions contributing to leading order decays}

Restricting ourselves again to glueball -meson interactions contributing to leading order decays
to two and four pions, the interaction Lagrangian
linear in $\D$ or $T$, up to quadratic in $\rho$, and maximally trilinear in $\pi,\rho$
reads, for the dilatonic mode,
\bea
\mathcal L_{\D}&=&\Tr \Biggl\{ d_1 
\frac{1}{2} \partial_\mu\pi\partial_\nu\pi 
\left(\eta^{\mu\nu}-\frac{\partial^\mu \partial^\nu}{\Box}\right)\D
+d_2 \MKK^2 \frac{1}{2} \rho_\mu \rho_\nu 
\left(\eta^{\mu\nu}-\frac{\partial^\mu \partial^\nu}{\Box}\right)\D
\nonumber\\
&&+d_3\frac{1}{2} \bar F_{\mu\rho}\bar F_\nu{}^\rho
\left(\eta^{\mu\nu}-\frac{\partial^\mu \partial^\nu}{\Box}\right)\D
+id_5\6_\mu\pi [\pi,\rho_\nu]
\left(\eta^{\mu\nu}-\frac{\partial^\mu \partial^\nu}{\Box}\right)\D
\Biggr\}
\eea
with coefficients
\bea\label{didef}
d_1&=&\int dZ \frac{H_D}{\pi K},
\qquad d_2=2\kappa\int dZ\,K(\psi_1')^2 H_D,\nonumber\\
\qquad d_3&=&2\kappa\int dZ\,K^{-1/3}(\psi_1)^2 H_D,
\qquad d_5=\int dZ \frac{\psi_1 H_D}{\pi K},
\eea
and for the tensor glueball
\bea\label{calLT}
\mathcal L_T&=&\Tr \biggl\{ 
\frac12 t_1\, \6_\mu\pi \6_\nu\pi \, T^{\mu\nu}
+\frac12 t_2 \,\MKK^2 \rho_\mu \rho_\nu \,T^{\mu\nu}\nonumber\\
&&\qquad +\frac12 t_3 \,\bar F_{\mu\rho}\bar F_{\nu}{}^{\rho}\, T^{\mu\nu}
+it_5\,\6_\mu\pi [\pi,\rho_\nu] \,T^{\mu\nu}
\biggr\}
\eea
with $t_i$ defined in analogy to (\ref{didef}). Because of $H_T\propto H_D$
with the normalization conditions (\ref{HDTnorm}), we simply have
$t_i=\sqrt6 d_i$.

The numerical results for $d_i$ and the corresponding coefficients $d_i^*$ for the
next-highest dilatonic scalar are given in Table \ref{tabct}.


\subsubsection{$\D$-4$\pi^0$ vertex from next-to-leading order DBI action}\label{sec:LD4pi0}

A direct coupling of glueball modes to more than two pions appears only at higher orders of the DBI action
of the D8 branes. For the coupling to four $\pi^0\equiv\pi^3$ we need to expand up to quartic terms in $F^3_{\nu Z}$.
The action, restricted to $F^3_{\nu Z}$, reads
\be
S=T_{\rm D8}(2\pi\alpha')^2\int d^9x e^{-\Phi}\sqrt{-\tilde g}
\left\{ -\frac12 g^{ZZ}g^{\mu\nu}F^3_{\mu Z}F^3_{\nu Z}
+\frac{(2\pi\alpha')^2}{8}\left[g^{ZZ}g^{\mu\nu}F^3_{\mu Z}F^3_{\nu Z}\right]^2 \right\}.
\ee

Inserting the metric fluctuations corresponding to the dilatonic glueball and
dropping terms that vanish on the mass shell of the glueball gives
\bea\label{LintGD4pi0}
\mathcal L^{\D\to4\pi^0}=3 d_1'\left[(\6_\mu\pi^0)^2\right]^2 \D
-2 d_1' (\6_\rho\pi^0)^2 (\6_\mu \pi^0)(\6_\nu \pi^0)\left(\eta^{\mu\nu}-\frac{\partial^\mu \partial^\nu}{M_D^2}\right)\D
\eea
with
\be
d_1'=\frac{3^9 \pi^3}{8 \lambda^3 N_c \MKK^4}
\int dZ\, H_D K^{-8/3}\approx 2.513\cdot 10^6 \,\lambda^{-\frac72}\, N_c^{-2} \MKK^{-5}.
\ee

\subsubsection{Two-glueball-two-$\pi$ vertices}\label{sec:LDD2pi0}

The leading (Yang-Mills) part of the DBI action also contains nonlinear terms with respect to
the metric fluctuations dual to glueballs, which have to be considered for the glueball
decays in four $\pi^0$, which vanish at leading order. 
Expanding the bilinear term in $\pi$ to second order in
the dilatonic mode $\D$ yields
\bea\label{LDD2pi0}
\mathcal L^{\D\D\pi\pi}=\frac{d_1^{DD}}{2}\Tr\biggl[&&
3(\6_\mu\pi)^2 \D^2 \nonumber\\
&&-\6_\mu\pi \6_\nu\pi
\,\eta_{\rho\sigma}
\left(\eta^{\mu\rho}-\frac{\partial^\mu \partial^\rho}{\Box}\right)\D
\left(\eta^{\sigma\nu}-\frac{\partial^\sigma \partial^\nu}{\Box}\right)\D
\biggr]
\eea
with
\be
d_1^{DD}=\int dZ \frac{H_D^2}{\pi K}=399.04\ldots \lambda^{-1}N_c^{-2}\MKK^{-2}.
\ee

In Eq.~(\ref{GDGE00}) we have considered for completeness also the decay
through the lowest exotic scalar glueball. For this process the relevant
terms in the interaction Lagrangian turn out to be
\bea\label{LGD2pi0}
\mathcal L^{\D\G\pi^0\pi^0}&=&c_1^{DE}\Tr\biggl[
\6_\mu\pi \6^\nu\pi \left(\frac{\6^\mu\6^\sigma}{M_E^2}\G\right)
\left(\eta_{\sigma\nu}-\frac{\partial_\sigma \partial_\nu}{\Box}\right)\D\nonumber\\
&&\qquad\quad -\frac14 \6_\mu\pi \6^\mu\pi
\left(\frac{\6^\rho\6^\sigma}{M_E^2}\G\right)
\left(\eta_{\sigma\rho}-\frac{\partial_\sigma \partial_\rho}{\Box}\right)\D
\biggr]
\eea
with
\be
c_1^{DE}=\int dZ \frac{H_D \bar H_E}{\pi K}=1653.9\ldots \lambda^{-1}N_c^{-2}\MKK^{-2}.
\ee

\section{Four-pion decay amplitudes and phase space integrals}\label{sec:phint}


\subsection{Decay of scalar glueballs into 4 massless pions involving $\pi^\pm$}

The leading-order decay amplitude of a glueball into four pions involves two pairs of pions with different isospin index
(thus excluding the case of four $\pi^0$'s).
If $\mathcal M$ is the amplitude for $G\to 2\pi^a 2\pi^b$ with fixed $a\not=b$, the total decay rate
of a glueball into 4 pions is given by
\be
\Gamma_{G\to4\pi}=\frac34\times\frac1{2M}\int d{\rm LIPS}_4(M) |\mathcal M|^2,
\ee
where the factor $\frac34$ is due to a factor of 3 for the three different pairs $a,b$ possible, and $\frac14$
is the symmetry factor for two pairs of identical particles.

For the decay of a particle at rest with mass $M$ into $n$ particles we have
\be
d{\rm LIPS}_n(M)=(2\pi)^4 \delta^4\left(M\delta^\mu_0-\sum_{A=1}^n p_A^\mu\right) \prod_{B=1}^n \frac{d^3p_B}{(2\pi)^3 2p_B^0}.
\ee

Useful details of how to organize the integration over the final momenta are given in Ref.~\cite{Hashimoto:2007ze}.
As a test of the numerical procedure for implementing these integrations we have used that for massless final
states the phase space integral with $\mathcal M\equiv 1$ can be done analytically with the result 
\cite{Byckling:1971vca}
\be
\int d{\rm LIPS}_n(M)=\frac{M^{2n-4}}{2(4\pi)^{2n-3}\Gamma(n)\Gamma(n-1)}.
\ee

\subsubsection{Four-pion decay amplitude for the dilatonic glueball}

Because the lowest glueball corresponding to an ``exotic'' polarization of the metric fluctuations has
a rather lengthy interaction Lagrangian and because we arrived at the conjecture that the next-higher scalar
(predominantly dilatonic) mode should be interpreted as the lowest scalar glueball of QCD, we shall give
the decay amplitude into 4 pions explicitly only for the latter.
Denoting the final pion four-momenta in $\D\to 2\pi^a 2\pi^b$ by $p,p',q,q'$ (see Fig.~\ref{fig:g4pi}) and defining
\be
a^\mu=q^\mu-p^\mu,\quad b^\mu=q'^\mu-p'^\mu,\quad r^\mu=p'^\mu+q'^\mu,\quad s^\mu=p^\mu+q^\mu,\quad r^0+s^0=M_D,
\ee
we find
\bea
i\mathcal M&=&\sqrt{2} g_{\rho\pi\pi}\left(\Delta_\rho(r)+\Delta_\rho(s)\right)d_5 \, \mathbf a\cdot \mathbf b\nonumber\\
&&+g_{\rho\pi\pi}^2\Delta_\rho(r)\Delta_\rho(s)\biggl\{
d_2\MKK^2 \mathbf a\cdot \mathbf b\nonumber\\
&&\qquad
+d_3 \Bigl[(a_0 b_0-\mathbf a\cdot \mathbf b)\,\mathbf r\cdot \mathbf s
+\mathbf a\cdot \mathbf b\,(r_0 s_0-\mathbf r\cdot \mathbf s)\nonumber\\
&&\qquad\qquad-(a_0 r_0-\mathbf a\cdot \mathbf r)\,\mathbf b\cdot \mathbf s
-(b_0 s_0-\mathbf b\cdot \mathbf s)\,\mathbf a\cdot \mathbf r\Bigr]
\biggr\}\nonumber\\
&&+\left(q\leftrightarrow q'\right),
\eea
where $\Delta_\rho(r)=1/(r_0^2-\mathbf r^2-m_\rho^2+im_\rho \Gamma_\rho)$ with $\Gamma_\rho$ given by
(\ref{Gammarho}).

\subsubsection{Scalar glueball decay through $\rho\pi\pi$ and $\rho\rho$}\label{sec:Grhopipi}
The use of a finite width of the $\rho$ meson in the propagator $\Delta_\rho$ corresponds to a partial resummation
of formally higher order diagrams. This seems to be natural in view of the fact that
$\Gamma_\rho/m_\rho$ is not very small, but it should be kept in mind that e.g.\ a correction of the residue
of the propagator is being dropped.

If the glueball decay were to be treated strictly perturbatively in inverse powers of $\lambda$ and $N_c$, 
one would neglect $\Gamma_\rho/m_\rho$ as a higher-order contribution and treat the $\rho$ meson as nearly stable.
Because to leading order there is no local vertex that would couple a glueball directly to four pions,
the leading-order process would then be given by a decay into on-shell $\rho\pi\pi$ with decay width
proportional to $\lambda^{-2}N_c^{-3}$ as long as the glueball mass is below the $2\rho$ threshold; glueballs with mass larger
than $2m_\rho$ would have the decay into two $\rho$ mesons as dominant process for the eventual
decay into four pions, with partial width
proportional to $\lambda^{-1}N_c^{-2}$.


We have evaluated the decay rates into $\rho\pi\pi$ (and $\rho\rho$ when $M>2m_\rho$)
as a cross-check of our results for the decay into
four pions,
which coincide in the limit of large $\lambda$,
\bea
&&\lim_{\lambda\to\infty} \lambda^2 \,\Gamma_{G\to4\pi}/M=
\lim_{\lambda\to\infty} \lambda^2 \,\Gamma_{G\to\rho\pi\pi}/M=\gamma_1 \quad \mbox{for $m_\rho<M<2 m_\rho$}, \\
&&\lim_{\lambda\to\infty} \lambda \,\Gamma_{G\to4\pi}/M=\lim_{\lambda\to\infty} \lambda \,\Gamma_{G\to2\rho}/M
=\gamma_2 \quad \mbox{for $M>2 m_\rho$},
\eea
with $\gamma_1\approx 22.074 N_c^{-3}$ for the glueball mode $\D$, and $\gamma_2\approx 6.451 N_c^{-2}$
when its mass is artificially raised to 1722 MeV.
Taking these strictly leading-order results as a basis for the decay width into four pions
would give somewhat higher numerical values than the above calculation involving a finite $\Gamma_\rho$.\footnote{%
Ref.~\cite{Hashimoto:2007ze} has added $\Gamma_{G\to4\pi}$ and $\Gamma_{G\to\rho\pi\pi}$
when comparing their results with experimental data, which we regard as overcounting.}

For the lowest (exotic) glueball mode, whose mass is not much higher than $m_\rho$, 
we find $\gamma_1\approx 0.0030
N_c^{-3}$ ($\approx 0.00114 N_c^{-3}$ if the contribution from the $\breve c$'s
is dropped). Again the results for $\Gamma_{G\to4\pi}/M$ converge to this limit for $\lambda\to\infty$. 
However,
for $\lambda=16.63$ the effect of resumming $\Gamma_\rho$ in the calculation of $\Gamma_{G\to4\pi}$
is now an increase of the decay width by more than two
orders of magnitude compared to the strictly perturbative result that corresponds to
a nearly stable $\rho$ meson with negligible width: the latter would give 
$\Gamma_{\G\to\rho\pi\pi\to4\pi}/M_E\approx 4.0\times 10^{-7}$
compared to $1.3\times 10^{-4}$ from $\Gamma_{\G\to4\pi}/M_E$ with resummed $\rho$ propagators.

The decay amplitudes for $G\to\rho\pi\pi$ are somewhat unwieldy, in particular for the exotic glueball mode.
We therefore give details only for the decay of the excited dilatonic glueball into two $\rho$ mesons.
(For the analogous decay of the lowest dilatonic glueball when its mass is raised above the $2\rho$
threshold is obtained by replacing $d_i^*$ by $d_i$.)

No phase space integration is involved in this process, but the polarizations of the $\rho$ meson
have to be summed over. Denoting the two transverse and the one longitudinal polarization by indices T
and L, respectively, the result is
\be
\Gamma_{D^*\to\rho\rho}/M_{D^*}=\frac{3}{32\pi M_{D^*}^3}
\sqrt{M_{D^*}^2-4m_\rho^2}\left(|\mathcal M_L|^2+2|\mathcal M_T|^2\right),
\ee
with 
\be
|\mathcal M_L|=\left|d_2^* \MKK^2 \frac{M_{D^*}^2}{4m_\rho^2}+d_3^* m_\rho^2 \right|,\quad
|\mathcal M_T|=\left|d_2^* \MKK^2 + d_3^*\left(\frac34 M_{D^*}^2 - 2 m_\rho^2\right)\right|.
\ee

Decays into two $\omega$ mesons, whose mass equals the $\rho$ meson mass
in the Sakai-Sugimoto model (in the real world it is only 1\% heavier),
is given by the same expression with 
the overall isospin multiplicity factor $3$ omitted.
(For the excited dilatonic glueball, also decay into two $\phi$ mesons becomes
relevant, though not for the lowest dilatonic glueball when its mass is raised
to one of the glueball candidates as the latter are all below the $2\phi$ threshold.)


\subsection{Decay of dilatonic glueball into 4 massless $\pi^0$}\label{sec:GDto4pi0}

With $\mathcal M$ the amplitude for $G_D\to 4\pi^0$, the total decay rate
is given by
\be
\Gamma_{G_D\to4\pi^0}=\frac{1}{24}\times\frac1{2M}\int d{\rm LIPS}_4(M) |\mathcal M|^2.
\ee
The dominant contribution is provided by (\ref{LintGD4pi0}), which leads to
\bea
\mathcal M/(8id_1')&=&3\left[p\cdot q\, p'\cdot q'+p\cdot p' \,q\cdot q'+p\cdot q' \,q\cdot p'\right]\nonumber\\
&&-p\cdot q \,\mathbf p'\cdot \mathbf q'-\mathbf p\cdot \mathbf q \,p'\cdot q'
-p\cdot p' \,\mathbf q\cdot \mathbf q'\nonumber\\
&&-\mathbf p\cdot \mathbf p' \,q\cdot q'
-p\cdot q' \,\mathbf q\cdot \mathbf p'-\mathbf p\cdot \mathbf q' \,q\cdot p'.
\eea

\subsection{Decay of tensor glueball into two vector mesons}\label{sec:MT2rho}

Unless one adjusts its mass parameter, only the excited tensor mode is above
the $2\rho$ threshold. The interaction Lagrangian (\ref{calLT}) with $t_i$ replaced
by $t_i^*$ determines
the amplitude in Eq.~(\ref{Tstorhorho}) for a specific tensor polarization $\epsilon_T^{\mu\nu}$ and
two $\rho$ mesons with momenta $p,q$ and polarizations $\epsilon^\mu(p,\lambda_1),\epsilon^\nu(q,\lambda_2)$ as
\bea
\mathcal M_{\epsilon_T}(\lambda_1,\lambda_2)&=&
\epsilon_\mu(p,\lambda_1)\epsilon_\nu(q,\lambda_2)\Bigl[t^*_2 \MKK^2 \epsilon_T^{\mu\nu} \nonumber\\
&&\quad
-t_3^* \left( p\cdot \epsilon_T\cdot q \,\eta^{\mu\nu}+p\cdot q \,\epsilon_T^{\mu\nu}
-p^\nu \epsilon_T^{\mu\rho}q_\rho-q^\mu \epsilon_T^{\nu\rho}p_\rho \right)\Bigr].
\eea
With $p^0=q^0=M/2$, $\mathbf p=-\mathbf q$, and $\mathbf p^2=(M/2)^2-\mr^2$, 
the result of the summation over the polarizations $\lambda_1$, $\lambda_2$ and the integration over
spatial directions reads
\bea
\int\frac{d\Omega}{4\pi}\sum_{\lambda_1,\lambda_2=1}^3 |\mathcal M_{\epsilon_T}(\lambda_1,\lambda_2)|^2&=&
2(t_2^* \MKK^2/\mr^2)^2 \left( \frac2{15} (\mathbf p^2)^2+\frac23 \mr^2 \,\mathbf p^2 + \mr^4 \right)\nonumber\\
&&+4 t_2^* t_3^* \MKK^2 \left(\frac{4}{3} \mathbf p^2 +\mr^2\right)\nonumber\\
&&+2t_3^{*2}\left(\frac{8}{15}(\mathbf p^2)^2+2 \mr^2 \,\mathbf p^2 + \mr^4
\right),
\eea
with $t_i^*\equiv\sqrt6 d_i^*$ and $d_i^*$ given in Table \ref{tabct}.

\bibliographystyle{kp}
\bibliography{glueballdecay}

\end{document}